# Testing hadronic-model predictions of depth of maximum of air-shower profiles and ground-particle signals using hybrid data of the Pierre Auger Observatory

A. Abdul Halim,[13] P. Abreu,[73] M. Aglietta,[55,53] I. Allekotte,[1] K. Almeida Cheminant,[71] A. Almela,[7,12] R. Aloisio,[46,47] J. Alvarez-Muñiz,[79] J. Ammerman Yebra,[79] G. A. Anastasi,[59,48] L. Anchordoqui,[86] B. Andrada,[7] S. Andringa,[73] L. Apollonio,[60,50] C. Aramo,[51] P. R. Araújo Ferreira,[43] E. Arnone,[64,53] J. C. Arteaga Velázquez,[68] P. Assis,[73] G. Avila,[11] E. Avocone,[58,47] A. Bakalova,[33] F. Barbato,[46,47] A. Bartz Mocellin,[85] J. A. Bellido,[13,70] C. Berat,[37] M. E. Bertaina,[64,53] G. Bhatta,[71] M. Bianciotto,[64,53] P. L. Biermann,[i] V. Binet,[5] K. Bismark,[40,7] T. Bister,[80,81] J. Biteau,[38,b] J. Blazek,[33] C. Bleve,[37] J. Blümer,[42] M. Boháčová,[33] D. Boncioli,[58,47] C. Bonifazi,[8,27] L. Bonneau Arbeletche,[22] N. Borodai,[71] J. Brack,[k] P. G. Brichetto Orchera,[7] F. L. Briechle,[43] A. Bueno,[78] S. Buitink,[15] M. Buscemi,[48,62] M. Büsken,[40,7] A. Bwembya,[80,81] K. S. Caballero-Mora,[67] S. Cabana-Freire,[79] L. Caccianiga,[60,50] F. Campuzano,[6] R. Caruso,[59,48] A. Castellina,[55,53] F. Catalani,[19] G. Cataldi,[49] L. Cazon,[79] M. Cerda,[10] A. Cermenati,[46,47] J. A. Chinellato,[22] J. Chudoba,[33] L. Chytka,[34] R. W. Clay,[13] A. C. Cobos Cerutti,[6] R. Colalillo,[61,51] M. R. Coluccia,[49] R. Conceição,[73] A. Condorelli,[38] G. Consolati,[50,56] M. Conte,[57,49] F. Convenga,[58,47] D. Correia dos Santos,[29] P. J. Costa,[73] C. E. Covault,[84] M. Cristinziani,[45] C. S. Cruz Sanchez,[3] S. Dasso,[4,2] K. Daumiller,[42] B. R. Dawson,[13] R. M. de Almeida,[29] J. de Jesús,[7,42] S. J. de Jong,[80,81] J. R. T. de Mello Neto,[27,28] I. De Mitri,[46,47] J. de Oliveira,[18] D. de Oliveira Franco,[49] F. de Palma,[57,49] V. de Souza,[20] B. P. de Souza de Errico,[27] E. De Vito,[57,49] A. Del Popolo,[59,48] O. Deligny,[35] N. Denner,[33] L. Deval,[42,7] A. di Matteo,[53] M. Dobre,[74] C. Dobrigkeit,[22] J. C. D'Olivo,[69] L. M. Domingues Mendes,[73,16] Q. Dorosti,[45] J. C. dos Anjos,[16] R. C. dos Anjos,[26] J. Ebr,[33] F. Ellwanger,[42] M. Emam,[80,81] R. Engel,[40,42] I. Epicoco,[57,49] M. Erdmann,[43] A. Etchegoyen,[7,12] C. Evoli,[46,47] H. Falcke,[80,82,81] G. Farrar,[88] A. C. Fauth,[22] N. Fazzini,[g] F. Feldbusch,[41] F. Fenu,[42,f] A. Fernandes,[73] B. Fick,[87] J. M. Figueira,[7] A. Filipčič,[77,76] T. Fitoussi,[42] B. Flaggs,[90] T. Fodran,[80] T. Fujii,[89,h] A. Fuster,[7,12] C. Galea,[80] C. Galelli,[60,50] B. García,[6] C. Gaudu,[39] H. Gemmeke,[41] F. Gesualdi,[7,42] A. Gherghel-Lascu,[74] P. L. Ghia,[35] U. Giaccari,[49] J. Glombitza,[43,d] F. Gobbi,[10] F. Gollan,[7] G. Golup,[1] M. Gómez Berisso,[1] P. F. Gómez Vitale,[11] J. P. Gongora,[11] J. M. González,[1] N. González,[7] D. Góra,[71] A. Gorgi,[55,53] M. Gottowik,[79] T. D. Grubb,[13] F. Guarino,[61,51] G. P. Guedes,[23] E. Guido,[45] L. Gülzow,[42] S. Hahn,[40] P. Hamal,[33] M. R. Hampel,[7] P. Hansen,[3] D. Harari,[1] V. M. Harvey,[13] A. Haungs,[42] T. Hebbeker,[43] C. Hojvat,[g] J. R. Hörandel,[80,81] P. Horvath,[34] M. Hrabovský,[34] T. Huege,[42,15] A. Insolia,[59,48] P. G. Isar,[75] P. Janecek,[33] V. Jilek,[33] J. A. Johnsen,[85] J. Jurysek,[33] K.-H. Kampert,[39] B. Keilhauer,[42] A. Khakurdikar,[80] V. V. Kizakke Covilakam,[7,42] H. O. Klages,[42] M. Kleifges,[41] F. Knapp,[40] J. Köhler,[42] N. Kunka,[41] B. L. Lago,[17] N. Langner,[43] M. A. Leigui de Oliveira,[25] Y. Lema-Capeans,[79] A. Letessier-Selvon,[36] I. Lhenry-Yvon,[35] L. Lopes,[73] L. Lu,[91] Q. Luce,[40] J. P. Lundquist,[76] A. Machado Payeras,[22] M. Majercakova,[33] D. Mandat,[33] B. C. Manning,[13] P. Mantsch,[g] F. M. Mariani,[60,50] A. G. Mariazzi,[3] I. C. Mariş,[14] G. Marsella,[62,48] D. Martello,[57,49] S. Martinelli,[42,7] O. Martínez Bravo,[65] M. A. Martins,[79] H.-J. Mathes,[42] J. Matthews,[a] G. Matthiae,[63,52] E. Mayotte,[85,39] S. Mayotte,[85] P. O. Mazur,[g] G. Medina-Tanco,[69] J. Meinert,[39] D. Melo,[7] A. Menshikov,[41] C. Merx,[42] S. Michal,[33] M. I. Micheletti,[5] L. Miramonti,[60,50] S. Mollerach,[1] F. Montanet,[37] L. Morejon,[39] C. Morello,[55,53] K. Mulrey,[80,81] R. Mussa,[53] W. M. Namasaka,[39] S. Negi,[33] L. Nellen,[69] K. Nguyen,[87] G. Nicora,[9] M. Niechciol,[45] D. Nitz,[87] D. Nosek,[32] V. Novotny,[32] L. Nožka,[34] A. Nucita,[57,49] L. A. Núñez,[31] C. Oliveira,[20] M. Palatka,[33] J. Pallotta,[9] S. Panja,[33] G. Parente,[79] T. Paulsen,[39] J. Pawlowsky,[39] M. Pech,[33] J. Pękala,[71] R. Pelayo,[66] L. A. S. Pereira,[24] E. E. Pereira Martins,[40,7] J. Perez Armand,[21] C. Pérez Bertolli,[7,42] L. Perrone,[57,49] S. Petrera,[46,47] C. Petrucci,[58,47] T. Pierog,[42] M. Pimenta,[73] M. Platino,[7] B. Pont,[80] M. Pothast,[81,80] M. Pourmohammad Shahvar,[62,48] P. Privitera,[89] M. Prouza,[33] S. Querchfeld,[39] J. Rautenberg,[39] D. Ravignani,[7] J. V. Reginatto Akim,[22] M. Reininghaus,[40] J. Ridky,[33] F. Riehn,[79] M. Risse,[45] V. Rizi,[58,47] W. Rodrigues de Carvalho,[80] E. Rodriguez,[7,42] J. Rodriguez Rojo,[11] M. J. Roncoroni,[7] S. Rossoni,[44] M. Roth,[42] E. Roulet,[1] A. C. Rovero,[4] P. Ruehl,[45] A. Saftoiu,[74] M. Saharan,[80] F. Salamida,[58,47] H. Salazar,[65] G. Salina,[52] J. D. Sanabria Gomez,[31] F. Sánchez,[7] E. M. Santos,[21] E. Santos,[33] F. Sarazin,[85] R. Sarmento,[73] R. Sato,[11] P. Savina,[91] C. M. Schäfer,[40] V. Scherini,[57,49] H. Schieler,[42] M. Schimassek,[35] M. Schimp,[39] D. Schmidt,[42] O. Scholten,[15,j] H. Schoorlemmer,[80,81] P. Schovánek,[33] F. G. Schröder,[90,42] J. Schulte,[43] T. Schulz,[42] S. J. Sciutto,[3] M. Scornavacche,[7,42] A. Sedoski,[7] A. Segreto,[54,48] S. Sehgal,[39] S. U. Shivashankara,[76] G. Sigl,[44] G. Silli,[7] O. Sima,[74,c] K. Simkova,[15] F. Simon,[41] R. Smau,[74] R. Šmída,[89] P. Sommers,[l] J. F. Soriano,[86] R. Squartini,[10] M. Stadelmaier,[50,60,42] S. Stanič,[76] J. Stasielak,[71] P. Stassi,[37] S. Strähnz,[40] M. Straub,[43] T. Suomijärvi,[38] A. D. Supanitsky,[7] Z. Svozilikova,[33] Z. Szadkowski,[72] F. Tairli,[13] A. Tapia,[30] C. Taricco,[64,53] C. Timmermans,[81,80] O. Tkachenko,[42] P. Tobiska,[33] C. J. Todero Peixoto,[19] B. Tomé,[73] Z. Torrès,[37] A. Travaini,[10] P. Travnicek,[33] C. Trimarelli,[58,47] M. Tueros,[3] M. Unger,[42] L. Vaclavek,[34] M. Vacula,[34] J. F. Valdés Galicia,[69] L. Valore,[61,51] E. Varela,[65] A. Vásquez-Ramírez,[31] D. Veberič,[42] C. Ventura,[28] I. D. Vergara Quispe,[3]

V. Verzi,[52] J. Vicha,[33] J. Vink,[83] S. Vorobiov,[76] C. Watanabe,[27] A. A. Watson,[e] A. Weindl,[42] L. Wiencke,[85] H. Wilczyński,[71] D. Wittkowski,[39] B. Wundheiler,[7] B. Yue,[39] A. Yushkov,[33] O. Zapparrata,[14] E. Zas,[79] D. Zavrtanik,[76,77] and M. Zavrtanik[77,76]

(The Pierre Auger Collaboration)[*]

[1]*Centro Atómico Bariloche and Instituto Balseiro (CNEA-UNCuyo-CONICET), San Carlos de Bariloche, Argentina*
[2]*Departamento de Física and Departamento de Ciencias de la Atmósfera y los Océanos, FCEyN, Universidad de Buenos Aires and CONICET, Buenos Aires, Argentina*
[3]*IFLP, Universidad Nacional de La Plata and CONICET, La Plata, Argentina*
[4]*Instituto de Astronomía y Física del Espacio (IAFE, CONICET-UBA), Buenos Aires, Argentina*
[5]*Instituto de Física de Rosario (IFIR)—CONICET/U.N.R. and Facultad de Ciencias Bioquímicas y Farmacéuticas U.N.R., Rosario, Argentina*
[6]*Instituto de Tecnologías en Detección y Astropartículas (CNEA, CONICET, UNSAM), and Universidad Tecnológica Nacional—Facultad Regional Mendoza (CONICET/CNEA), Mendoza, Argentina*
[7]*Instituto de Tecnologías en Detección y Astropartículas (CNEA, CONICET, UNSAM), Buenos Aires, Argentina*
[8]*International Center of Advanced Studies and Instituto de Ciencias Físicas, ECyT-UNSAM and CONICET, Campus Miguelete—San Martín, Buenos Aires, Argentina*
[9]*Laboratorio Atmósfera—Departamento de Investigaciones en Láseres y sus Aplicaciones—UNIDEF (CITEDEF-CONICET), Argentina*
[10]*Observatorio Pierre Auger, Malargüe, Argentina*
[11]*Observatorio Pierre Auger and Comisión Nacional de Energía Atómica, Malargüe, Argentina*
[12]*Universidad Tecnológica Nacional—Facultad Regional Buenos Aires, Buenos Aires, Argentina*
[13]*University of Adelaide, Adelaide, S.A., Australia*
[14]*Université Libre de Bruxelles (ULB), Brussels, Belgium*
[15]*Vrije Universiteit Brussels, Brussels, Belgium*
[16]*Centro Brasileiro de Pesquisas Fisicas, Rio de Janeiro, RJ, Brazil*
[17]*Centro Federal de Educação Tecnológica Celso Suckow da Fonseca, Petropolis, Brazil*
[18]*Instituto Federal de Educação, Ciência e Tecnologia do Rio de Janeiro (IFRJ), Brazil*
[19]*Universidade de São Paulo, Escola de Engenharia de Lorena, Lorena, SP, Brazil*
[20]*Universidade de São Paulo, Instituto de Física de São Carlos, São Carlos, SP, Brazil*
[21]*Universidade de São Paulo, Instituto de Física, São Paulo, SP, Brazil*
[22]*Universidade Estadual de Campinas (UNICAMP), IFGW, Campinas, SP, Brazil*
[23]*Universidade Estadual de Feira de Santana, Feira de Santana, Brazil*
[24]*Universidade Federal de Campina Grande, Centro de Ciencias e Tecnologia, Campina Grande, Brazil*
[25]*Universidade Federal do ABC, Santo André, SP, Brazil*
[26]*Universidade Federal do Paraná, Setor Palotina, Palotina, Brazil*
[27]*Universidade Federal do Rio de Janeiro, Instituto de Física, Rio de Janeiro, RJ, Brazil*
[28]*Universidade Federal do Rio de Janeiro (UFRJ), Observatório do Valongo, Rio de Janeiro, RJ, Brazil*
[29]*Universidade Federal Fluminense, EEIMVR, Volta Redonda, RJ, Brazil*
[30]*Universidad de Medellín, Medellín, Colombia*
[31]*Universidad Industrial de Santander, Bucaramanga, Colombia*
[32]*Charles University, Faculty of Mathematics and Physics, Institute of Particle and Nuclear Physics, Prague, Czech Republic*
[33]*Institute of Physics of the Czech Academy of Sciences, Prague, Czech Republic*
[34]*Palacky University, Olomouc, Czech Republic*
[35]*CNRS/IN2P3, IJCLab, Université Paris-Saclay, Orsay, France*
[36]*Laboratoire de Physique Nucléaire et de Hautes Energies (LPNHE), Sorbonne Université, Université de Paris, CNRS-IN2P3, Paris, France*
[37]*Univ. Grenoble Alpes, CNRS, Grenoble Institute of Engineering Univ. Grenoble Alpes, LPSC-IN2P3, 38000 Grenoble, France*
[38]*Université Paris-Saclay, CNRS/IN2P3, IJCLab, Orsay, France*
[39]*Bergische Universität Wuppertal, Department of Physics, Wuppertal, Germany*
[40]*Karlsruhe Institute of Technology (KIT), Institute for Experimental Particle Physics, Karlsruhe, Germany*

[41]*Karlsruhe Institute of Technology (KIT), Institut für Prozessdatenverarbeitung und Elektronik, Karlsruhe, Germany*
[42]*Karlsruhe Institute of Technology (KIT), Institute for Astroparticle Physics, Karlsruhe, Germany*
[43]*RWTH Aachen University, III. Physikalisches Institut A, Aachen, Germany*
[44]*Universität Hamburg, II. Institut für Theoretische Physik, Hamburg, Germany*
[45]*Universität Siegen, Department Physik—Experimentelle Teilchenphysik, Siegen, Germany*
[46]*Gran Sasso Science Institute, L'Aquila, Italy*
[47]*INFN Laboratori Nazionali del Gran Sasso, Assergi (L'Aquila), Italy*
[48]*INFN, Sezione di Catania, Catania, Italy*
[49]*INFN, Sezione di Lecce, Lecce, Italy*
[50]*INFN, Sezione di Milano, Milano, Italy*
[51]*INFN, Sezione di Napoli, Napoli, Italy*
[52]*INFN, Sezione di Roma "Tor Vergata", Roma, Italy*
[53]*INFN, Sezione di Torino, Torino, Italy*
[54]*Istituto di Astrofisica Spaziale e Fisica Cosmica di Palermo (INAF), Palermo, Italy*
[55]*Osservatorio Astrofisico di Torino (INAF), Torino, Italy*
[56]*Politecnico di Milano, Dipartimento di Scienze e Tecnologie Aerospaziali, Milano, Italy*
[57]*Università del Salento, Dipartimento di Matematica e Fisica "E. De Giorgi", Lecce, Italy*
[58]*Università dell'Aquila, Dipartimento di Scienze Fisiche e Chimiche, L'Aquila, Italy*
[59]*Università di Catania, Dipartimento di Fisica e Astronomia "Ettore Majorana", Catania, Italy*
[60]*Università di Milano, Dipartimento di Fisica, Milano, Italy*
[61]*Università di Napoli "Federico II", Dipartimento di Fisica "Ettore Pancini", Napoli, Italy*
[62]*Università di Palermo, Dipartimento di Fisica e Chimica "E. Segrè", Palermo, Italy*
[63]*Università di Roma "Tor Vergata", Dipartimento di Fisica, Roma, Italy*
[64]*Università Torino, Dipartimento di Fisica, Torino, Italy*
[65]*Benemérita Universidad Autónoma de Puebla, Puebla, México*
[66]*Unidad Profesional Interdisciplinaria en Ingeniería y Tecnologías Avanzadas del Instituto Politécnico Nacional (UPIITA-IPN), México, D.F., México*
[67]*Universidad Autónoma de Chiapas, Tuxtla Gutiérrez, Chiapas, México*
[68]*Universidad Michoacana de San Nicolás de Hidalgo, Morelia, Michoacán, México*
[69]*Universidad Nacional Autónoma de México, México, D.F., México*
[70]*Universidad Nacional de San Agustin de Arequipa, Facultad de Ciencias Naturales y Formales, Arequipa, Peru*
[71]*Institute of Nuclear Physics PAN, Krakow, Poland*
[72]*University of Łódź, Faculty of High-Energy Astrophysics,Łódź, Poland*
[73]*Laboratório de Instrumentação e Física Experimental de Partículas—LIP and Instituto Superior Técnico—IST, Universidade de Lisboa—UL, Lisboa, Portugal*
[74]*"Horia Hulubei" National Institute for Physics and Nuclear Engineering, Bucharest-Magurele, Romania*
[75]*Institute of Space Science, Bucharest-Magurele, Romania*
[76]*Center for Astrophysics and Cosmology (CAC), University of Nova Gorica, Nova Gorica, Slovenia*
[77]*Experimental Particle Physics Department, J. Stefan Institute, Ljubljana, Slovenia*
[78]*Universidad de Granada and C.A.F.P.E., Granada, Spain*
[79]*Instituto Galego de Física de Altas Enerxías (IGFAE), Universidade de Santiago de Compostela, Santiago de Compostela, Spain*
[80]*IMAPP, Radboud University Nijmegen, Nijmegen, The Netherlands*
[81]*Nationaal Instituut voor Kernfysica en Hoge Energie Fysica (NIKHEF), Science Park, Amsterdam, The Netherlands*
[82]*Stichting Astronomisch Onderzoek in Nederland (ASTRON), Dwingeloo, The Netherlands*
[83]*Universiteit van Amsterdam, Faculty of Science, Amsterdam, The Netherlands*
[84]*Case Western Reserve University, Cleveland, Ohio, USA*
[85]*Colorado School of Mines, Golden, Colorado, USA*
[86]*Department of Physics and Astronomy, Lehman College, City University of New York, Bronx, New York, USA*
[87]*Michigan Technological University, Houghton, Michigan, USA*
[88]*New York University, New York, New York, USA*
[89]*University of Chicago, Enrico Fermi Institute, Chicago, Illinois, USA*

[90]*University of Delaware, Department of Physics and Astronomy, Bartol Research Institute, Newark, Delaware, USA*
[91]*University of Wisconsin-Madison, Department of Physics and WIPAC, Madison, Wisconsin, USA*

We test the predictions of hadronic interaction models regarding the depth of maximum of air-shower profiles, $X_{\max}$, and ground-particle signals in water-Cherenkov detectors at 1000 m from the shower core, $S(1000)$, using the data from the fluorescence and surface detectors of the Pierre Auger Observatory. The test consists of fitting the measured two-dimensional ($S(1000)$, $X_{\max}$) distributions using templates for simulated air showers produced with hadronic interaction models Epos-LHC, QGSJet-II-04, SIBYLL 2.3d and leaving the scales of predicted $X_{\max}$ and the signals from hadronic component at ground as free-fit parameters. The method relies on the assumption that the mass composition remains the same at all zenith angles, while the longitudinal shower development and attenuation of ground signal depend on the mass composition in a correlated way. The analysis was applied to 2239 events detected by both the fluorescence and surface detectors of the Pierre Auger Observatory with energies between $10^{18.5}$ eV to $10^{19.0}$ eV and zenith angles below 60°. We found, that within the assumptions of the method, the best description of the data is achieved if the predictions of the hadronic interaction models are shifted to deeper $X_{\max}$ values and larger hadronic signals at all zenith angles. Given the magnitude of the shifts and the data sample size, the statistical significance of the improvement of data description using the modifications considered in the paper is larger than $5\sigma$ even for any linear combination of experimental systematic uncertainties.

## I. INTRODUCTION

The dominant contribution to uncertainties in the determination of the mass composition of ultrahigh-energy cosmic rays (UHECR, energy $E > 10^{18.0}$ eV) comes from the modeling of extensive air showers. Modern hadronic interaction models used for this purpose are based on extrapolation of interaction parameters like cross sections, multiplicities, elasticities, etc., measured at accelerators at lower beam energies up to $\sqrt{s} = 13$ TeV for proton-proton ($pp$) collisions at the LHC and pseudorapidities $|\eta| \lesssim 5$ compared to energies $\sqrt{s} \gtrsim 50$ TeV and pseudorapidities[1] $\eta \approx 7$ to 11 driving the energy flow of the first interactions of UHECR in the atmosphere, where the target is different (mostly oxygen and nitrogen nuclei). Therefore, improvements in the description of the LHC data, implemented in the modern models do not necessarily lead to unambiguous, nearly hadronic model-independent predictions for the mass-sensitive air-shower observables. For instance, at $10^{18.7}$ eV the span in predictions for the mean depth of maximum of air shower profiles, $\langle X_{\max}\rangle$, between models used by the UHECR community (Epos-LHC [1], QGSJet-II-04 [2], SIBYLL 2.3d [3]) is $\sim 25$ g/cm$^2$, nearly independently of the primary particle mass and energy. Such a difference can be considered only a lower limit on the systematic uncertainty of the predicted $X_{\max}$ scale. This is about one-quarter of the difference between $\langle X_{\max}\rangle$ values of the two astrophysical extremes, protons and iron nuclei. As a consequence, the mass composition of cosmic rays can be referred only with respect to $X_{\max}$ scale predicted by a particular model. The largest differences in the predicted $\langle X_{\max}\rangle$ between the models, with a minimal impact on the elongation rate, come from the properties of the first hadronic interaction and production of nucleons-antinucleons in pion-air and kaon-air interactions that are not well-known in the relevant kinematic region (for more detailed discussion, see e.g., Ref. [4]). The general comparison of properties and different treatments of hadronic interactions by the three models used in this work is summarized in Ref. [5]. Note that although the models used in this work treat differently the properties of hadronic interactions, the range of air-shower properties predicted by these models does not need to include all the possibilities.

[*]auger_spokespersons@fnal.gov
The Pierre Auger Observatory, Av. San Martín Norte 306, 5613 Malargüe, Mendoza, Argentina.
[a]Louisiana State University, Baton Rouge, Louisiana, USA.
[b]Institut universitaire de France (IUF), France.
[c]Also at University of Bucharest, Physics Department, Bucharest, Romania.
[d]Now at ECAP, Erlangen, Germany.
[e]School of Physics and Astronomy, University of Leeds, Leeds, United Kingdom.
[f]Now at Agenzia Spaziale Italiana (ASI). Via del Politecnico 00133, Roma, Italy.
[g]Fermi National Accelerator Laboratory, Fermilab, Batavia, Illinois, USA.
[h]Now at Graduate School of Science, Osaka Metropolitan University, Osaka, Japan.
[i]Max-Planck-Institut für Radioastronomie, Bonn, Germany.
[j]Also at Kapteyn Institute, University of Groningen, Groningen, The Netherlands.
[k]Colorado State University, Fort Collins, Colorado, USA.
[l]Pennsylvania State University, University Park, Pennsylvania, USA.

[1]The forward calorimeters at LHC can cover $\eta = 8.4$ to 15, but of neutral particles only.

The difference in the standard deviation of $X_{\max}$ distributions, $\sigma(X_{\max})$, between the models is within $\sim 5\ \mathrm{g/cm^2}$, whereas the difference between the $X_{\max}$ fluctuations of protons and iron nuclei is $\sim 40\ \mathrm{g/cm^2}$. Therefore, the difference of $\sigma(X_{\max})$ in model predictions has a smaller effect on the mass composition inferences compared to the difference in predictions of $\langle X_{\max}\rangle$. Note, that there is no direct correspondence of the model scales of $X_{\max}$ and $\sigma(X_{\max})$, i.e., the differences in $\sigma(X_{\max})$ are not a mere consequence of the differences in $\langle X_{\max}\rangle$.

In general, the signal produced by air-shower particles reaching the ground shows much lower sensitivity to the mass composition than in the case of $X_{\max}$. The model differences in predictions of the ground-particle signal, for instance, at 1000 m from the impact point of an air shower at $10^{18.7}$ eV detected at the Pierre Auger Observatory (Auger) [6] are at the level of $\sim 3$ VEM,[2] whereas the difference between protons and iron nuclei is about twice this value. The fluctuations of the ground signal are dominated by the detector resolution, suppressing any significant sensitivity to the model or primary mass [7].

### A. Problems in the description of data with hadronic interaction models

The correctness of the model predictions can be tested using data from air-shower experiments. At the Pierre Auger Observatory, for instance, negative variance of the logarithm of primary masses [$\sigma^2(\ln A)$] and poor description of the measured $X_{\max}$ distributions are obtained when simulations with QGSJet-II-04 are used for the interpretation of the measured $X_{\max}$ moments in terms of the moments of $\ln A$ for $E \gtrsim 10^{18.5}$ eV [8–11]. The problem is that, due to relatively shallow $\langle X_{\max}\rangle$ predictions of QGSJet-II-04, the best possible description of the data is achieved with proton-helium mixes, but for these mixes the modeled $X_{\max}$ distributions are broader than the observed ones. These findings are further supported by the observation of the negative correlation between $X_{\max}$ and the signal in surface detector (SD) [11,12] in the Auger data near the "ankle" feature ($E \approx 10^{18.7}$ eV) of the UHECR spectrum. That can be achieved only for the mixed primary composition containing particles heavier than helium nuclei, which is a robust statement based on the general phenomenology of air-shower development and, therefore, not sensitive to the properties of a particular hadronic model.

The deficiencies of the models are more evident for the SD observables, where, in some cases, the data are not even bracketed by the Monte Carlo (MC) predictions for protons and iron nuclei. The muon deficit in simulations, known as the "muon puzzle" [5], is the best-known example of this kind. In particular, this was observed by Auger for inclined showers (zenith angles $\theta = 62°$ to $80°$) dominated by the muon component, for Epos-LHC and QGSJet-II-04 at $\sim 10^{19.0}$ eV [13], and in direct measurements with underground muon detectors at $E = 10^{17.3}$ eV to $10^{18.3}$ eV and $\theta < 45°$ [14]. At the same time, the fluctuations of the muon signal measured by Auger [15] are consistent with the MC predictions, indicating that the muon deficit might originate from the accumulation of small deviations from the model predictions during the development of a shower, rather than be caused by a strong deviation in the first interaction. The range of predictions for the muon production depth of protons and iron nuclei is outside of the measured values for Epos-LHC above $10^{19.3}$ eV [16].

Less directly, the muon deficit in simulations was observed as a deficit of the total signal at 1000 m from the shower core, $S(1000)$, in the Auger SD stations [17] for vertical ($\theta < 60°$) showers with energies around $10^{19}$ eV. In this analysis, within the assumption that the electromagnetic (em) component, $X_{\max}$, and, correspondingly, the mass composition inferred from the $X_{\max}$ measurements, are predicted correctly by a particular model, the deficit of $S(1000)$ was interpreted as the underestimation of the hadronic signal (dominated by muons) in simulations by $(33 \pm 16)\%$ for Epos-LHC and by $(61 \pm 21)\%$ for QGSJet-II-04, with a strong dependence on the energy scale. An indicative summary of all these tests for the three models used in this work is given in Table I.

The use of an incorrect MC $X_{\max}$ scale would lead one to a biased inference on the mass composition and, through this, to a biased estimate of the muon deficit, since the muon content in a shower scales with the primary mass $\propto A^{1-\beta}$, $\beta \approx 0.9$ [18]. Therefore, in a more comprehensive approach a modification of the MC scales of both $X_{\max}$ and SD signals, going along with the fitting of the primary mass composition accounting for these modifications, should be considered.

### B. Progressive testing of hadronic interaction models

In this work, progressive testing of the model predictions is performed. First, we allow for a rescaling of the signal on the ground produced by the hadronic shower component at 1000 m with a factor $R_{\mathrm{had}}$. Then we add a zenith-angle dependence of $R_{\mathrm{had}}$ defining two parameters $R_{\mathrm{had}}(\theta_{\min})$ and $R_{\mathrm{had}}(\theta_{\max})$ at the two extreme zenith-angle bins. We assume a linear dependence of $R_{\mathrm{had}}$ on the distance of $X_{max}$ to the ground in atmospheric depth units to relate $R_{had}$ to different zenith angles. Finally, we consider a shift in the predicted $X_{max}$ distributions ($\Delta X_{max}$) that is assumed to be independent of the primary mass and energy. In this way, we consider freedom not only in the scale of the simulated hadronic part of the ground signal but also in the simulated $X_{max}$ scale. Consequently, the main differences in $X_{max}$ and $S(1000)$ predictions of the models are reduced, and similar mass composition inferences for the Auger data are obtained. It is remarkable, that a consistent description of the Auger data with all three models can be achieved with the modification of only two scale parameters.

[2]This unit is the signal produced by a muon traversing the station on a vertical trajectory.

TABLE I. Indicative summary of the results of tests of models using Auger data (✓—no tension, ✗—tension). In the case of SIBYLL 2.3d, we also show estimations based on the previous version of the model Sibyll when available in the literature.

| Test | Energy/EeV | $\theta/°$ | Epos-LHC | QGSJet-II-04 | SIBYLL 2.3d |
|---|---|---|---|---|---|
| $X_{max}$ moments [8–11] | ~3 to 50 | 0 to 80 | ✓ | ✗ | ✓ (2.3c) |
| $X_{max}$: $S(1000)$ correlation [11,12] | 3 to 10 | 0 to 60 | ✓ | ✗ | ✓ (2.3c) |
| Mean muon number [13,15] | ~10 | ~67 | ✗ | ✗ | ✗ |
| Mean muon number [14] | 0.2 to 2 | 0 to 45 | ✗ | ✗ | ... |
| Fluctuation of muon number [15] | 4 to 40 | ~67 | ✓ | ✓ | ✓ |
| Muon production depth [16] | 20 to 70 | ~60 | ✗ | ✓ | ... |
| $S(1000)$ [17] | ~10 | 0 to 60 | ✗ | ✗ | ... |

The analysis is performed for the energy region around the ankle ($E \approx 10^{18.7}$ eV) in the energy spectrum where the UHECR mass composition is mixed [9,11,12]. Specifically, we find the values of $\Delta X_{max}$, $R_{had}(\theta)$, and fractions of four primary particles (protons, and helium, oxygen, iron nuclei) for which the best-fit of the measured two-dimensional distributions of $(S(1000), X_{max})$ is achieved. The remaining differences between the predictions of models with a smaller effect on MC templates like the fluctuations of $X_{max}$ and hadronic signal, and the mass composition dependence of $R_{had}(\theta)$ and $\Delta X_{max}$ are not considered in this work.

In Sec. II, we give a detailed description of the method. In Sec. III, the method is applied to the data of the Pierre Auger Observatory. The results of the data analysis are discussed in Sec. IV followed by a summary of our findings in the Conclusions.

## II. METHOD

The method stems from Ref. [19] and it was first introduced in Ref. [20] followed by a slight modification in the approach to the ground-signal rescaling. A preliminary application of the method to the Auger data was presented in Ref. [21].

We perform a binned maximum-likelihood fit of the measured two-dimensional distributions $(S, X)$ with MC templates for four primary species simultaneously in five zenith-angle bins. The observables are $S(1000)$ and $X_{max}$ corrected for the energy evolution using the fluorescence detector (FD) energy $E_{FD}$,

$$S = S(1000)\left(\frac{E^{ref}}{E_{FD}}\right)^{1/B}, \qquad (1)$$

and

$$X = X_{max} + D\,\lg\left(\frac{E^{ref}}{E_{FD}}\right), \qquad (2)$$

where $B = 1.031 \pm 0.004$ is the SD energy calibration parameter [22] and the elongation rate of a single primary $D = 58$ g/cm$^2$/decade is taken as the average value over the four primary particles and three models used in this work. The value of the elongation rate varies only within $\pm 2$ g/cm$^2$ around the mean, in accordance with the universality with respect to the primary mass predicted within the phenomenological model in Ref. [23]. This is a consequence of energy dependencies of multiplicity, elasticity, and cross section assumed in this phenomenological model. We chose the reference energy $E^{ref} = 10^{18.7}$ eV for the analyzed energy range of $E_{FD} = 10^{18.5}$ eV to $10^{19.0}$ eV.

The signal $S(1000)$ is assumed to be composed of the hadronic ($S_{had}$) and electromagnetic ($S_{em}$) components. The signal $S_{had}$ is produced by muons, em particles from muon decays and low-energy neutral pions as in [17] according to the four-component shower universality model [24,25]. The signal $S_{em}$ is produced by em particles originating from high-energy neutral pions.

### A. Monte Carlo templates

The MC templates were prepared using simulated showers from a library produced within the Pierre Auger Collaboration [26]. Air showers were generated with CORSIKA 7.7400 [27–29] with a flat zenith-angle distribution in $\sin^2\theta$ [for $\theta \in (0°, 65°)$] and the energy distribution $\propto E^{-1}$. Subsequently, we reweight the events to match the measured energy spectrum [22]. Four different atmospheric profiles are considered to represent the typical variations of atmospheric conditions at the location of the Pierre Auger Observatory. The simulation set includes three models (Epos-LHC, QGSJet-II-04, SIBYLL 2.3d) and four primary particles (p, He, O, Fe). In productions with Epos-LHC and SIBYLL 2.3d, the low-energy ($E_{\mathrm{kin}} < 80$ GeV) interactions were simulated with URQMD [30], while QGSJet-II-04 was used in combination with FLUKA [31,32]. No significant dependency of the results on the choice of the low-energy model was found.

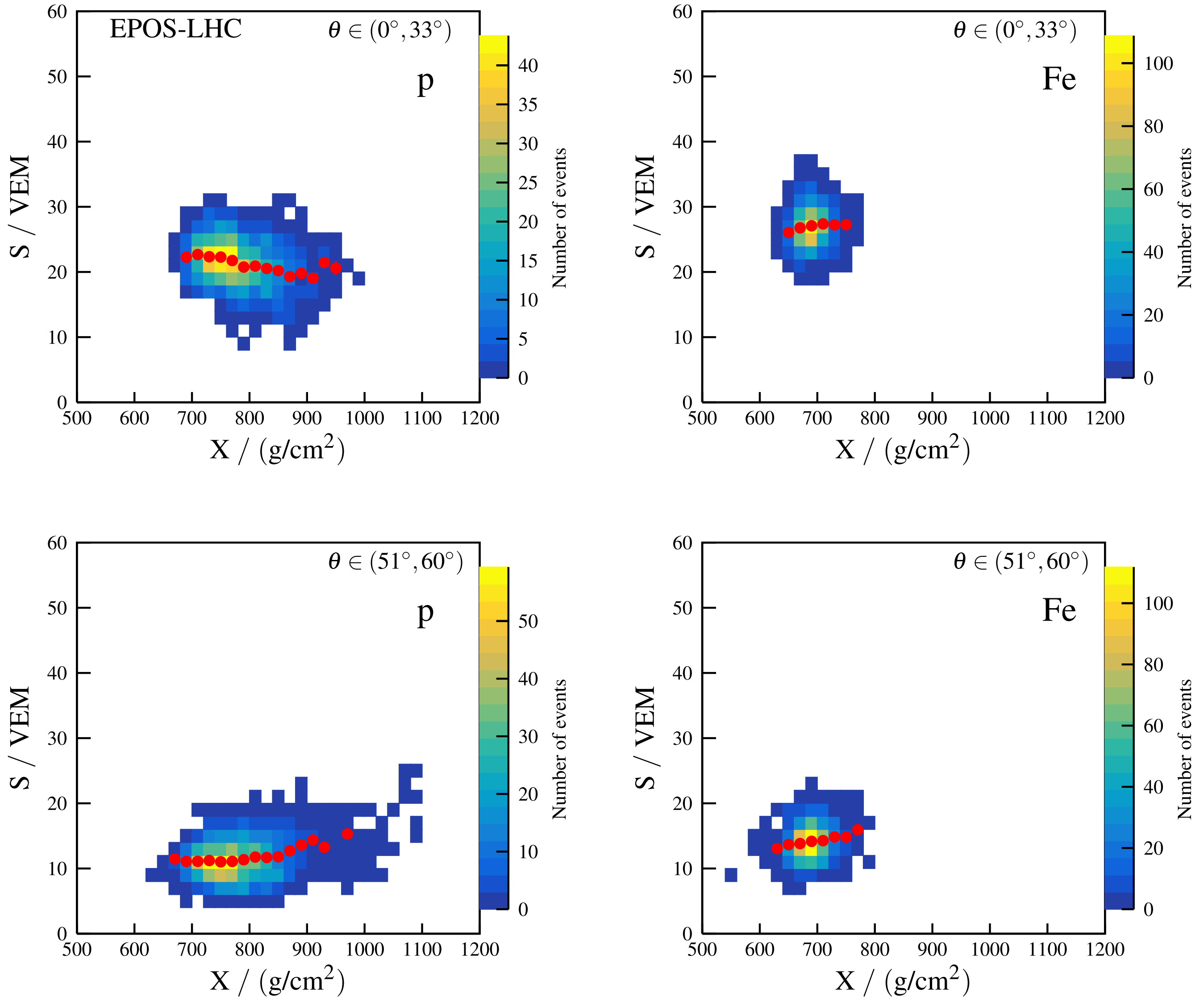

FIG. 1. Examples of two-dimensional distributions of $S$ and $X$ for protons (left) and iron nuclei (right) generated with Epos-LHC for zenith angles between 0° and 33° (top) and 51° and 60° (bottom). The red points indicate the mean values of $S$. $E = 10^{18.5}$ eV to $10^{19.0}$ eV.

The detector simulations and event reconstruction were performed with the Auger Offline Framework [33]. In the standard event processing chain, not all effects related to the detector calibration, atmospheric conditions, long-term performances, etc., are included in the simulations and reconstruction. To account for them, for the FD part an additional smearing of $X_{\mathrm{max}}$ distributions by ≈9 g/cm$^2$ is applied [9]. In the case of the SD part, the smearing of $S(1000)$ by 9%, corresponding to the maximal expected contribution from realistic operational conditions, was tested without finding any statistically significant effect on the results. The event selection is the same as applied to the data (see Sec. III). After the selection, the MC templates contain ≈15000 showers per primary species and model.

The analysis was carried out by splitting these simulated air-showers in five zenith-angle ranges containing nearly the same number of events, namely (0°, 33°), (33°, 39°), (39°, 45°), (45°, 51°), (51°, 60°). Examples of the $(S, X)$ distributions for protons and iron nuclei in the most vertical and most inclined angular ranges are shown in Fig. 1. Such two-dimensional distributions are then normalized and fitted with the ansatz function $\Phi$ described in detail in Appendix A. This function is a convolution of the generalized Gumbel distribution of $X$ and the Gaussian distribution of $S$ with the mean value linearly changing with $X$, reflecting in this way their correlation indicated in Fig. 1. A set of these trial functions for each model, primary particle, and zenith-angle range is used as MC templates in the following fitting procedure.

### B. Fitting procedure

For each model, we search for the most likely combination of the composition mix of the four primary species, the zenith-dependent rescaling parameter of the hadronic signal $R_{\mathrm{had}}(\theta)$, and the constant shift of the depth of shower maximum $\Delta X_{\max}$ in all the MC templates. The fitting method is a generalization of the fitting procedure used in Ref. [10] in the case of the $X_{\max}$ distribution and applied here to the two-dimensional $(X, S)$ distributions in five zenith-angle ranges simultaneously.

The negative log-likelihood-ratio expression that is minimized for a given model is of the form,

$$\ln\mathcal{L} = \begin{cases} \sum_k \sum_j \left(C_{jk} - n_{jk} + n_{jk}\ln\frac{n_{jk}}{C_{jk}}\right), & n_{jk} > 0, \\ \sum_k \sum_j C_{jk}, & n_{jk} = 0, \end{cases} \tag{3}$$

with the sums running over the two-dimensional bins $j$ in $(X, S)$ for the five $\theta$-bins $k$. The corresponding number of showers measured in bins $j$, $k$ is denoted by $n_{jk}$ and the predicted number of MC showers by $C_{jk}$. The latter number is obtained using the total number of measured showers $N^k_{\mathrm{data}}$ as

$$C_{jk} = N^k_{\mathrm{data}} \sum_i f_i \Phi_{i,k}(X'_{\mathrm{j,k}}, S'_{\mathrm{j,k}}), \tag{4}$$

where $\Phi_{i,k}$ denotes the template function $\Phi$ in $\theta$-bin $k$ for a given model and primary particle $i$ with relative fraction $f_i$. The modified $X$ prediction is of the form

$$X'_{\mathrm{j,k}} = X_{\mathrm{j,k}} + \Delta X_{\max} \tag{5}$$

and the rescaled predicted ground signal is

$$S'_{\mathrm{j,k}} = S_{\mathrm{j,k}} f^k_{\mathrm{SD}}(R_{\mathrm{had}}(\theta), \Delta X_{\max}), \tag{6}$$

where $X_{\mathrm{j,k}}$ and $S_{\mathrm{j,k}}$ are the center bin values of $X$ and $S$, respectively, of the original MC distribution $(X, S)$.

The rescaling parameter $f^k_{\mathrm{SD}}$ of all signals $S_{\mathrm{j,k}}$ is calculated as

$$f^k_{\mathrm{SD}} = R_{\mathrm{had}}(\theta)\frac{(E^{\mathrm{ref}})^{\beta-1/B}}{\langle E_{\mathrm{FD}}^{\beta-1/B}\rangle_k} g_{\mathrm{had},k} f_{\mathrm{had},k} + \frac{(E^{\mathrm{ref}})^{1-1/B}}{\langle E_{\mathrm{FD}}^{1-1/B}\rangle_k} g_{\mathrm{em},k}(1 - f_{\mathrm{had},k}), \tag{7}$$

where $S(1000) \propto E^B_{\mathrm{FD}}$ is assumed to be composed only of $S_{\mathrm{had}} \propto E^\beta_{\mathrm{FD}}$ and $S_{\mathrm{em}} \propto E_{\mathrm{FD}}$, and $f_{\mathrm{had}} = S_{\mathrm{had}}/S(1000)$. The parameter $\beta = 0.92$ is chosen following Ref. [23]. The mean energy factors $\langle E_{\mathrm{FD}}^{\beta-1/B}\rangle_k$ and $\langle E_{\mathrm{FD}}^{1-1/B}\rangle_k$ are calculated from all measured showers in the energy range $10^{18.5}$ eV to $10^{19.0}$ eV and $\theta$-bin $k$.[3] The mean hadronic fraction, see also Fig. 3, in the $\theta$-bin $k$, $f_{\mathrm{had},k}$, is calculated using the average hadronic fractions $f_{\mathrm{had},k,i}$ for simulated showers induced by a primary $i$ and weighted over the relative primary fractions $f_i$ as

$$f_{\mathrm{had},k} = \sum_i f_i f_{\mathrm{had},k,i}. \tag{8}$$

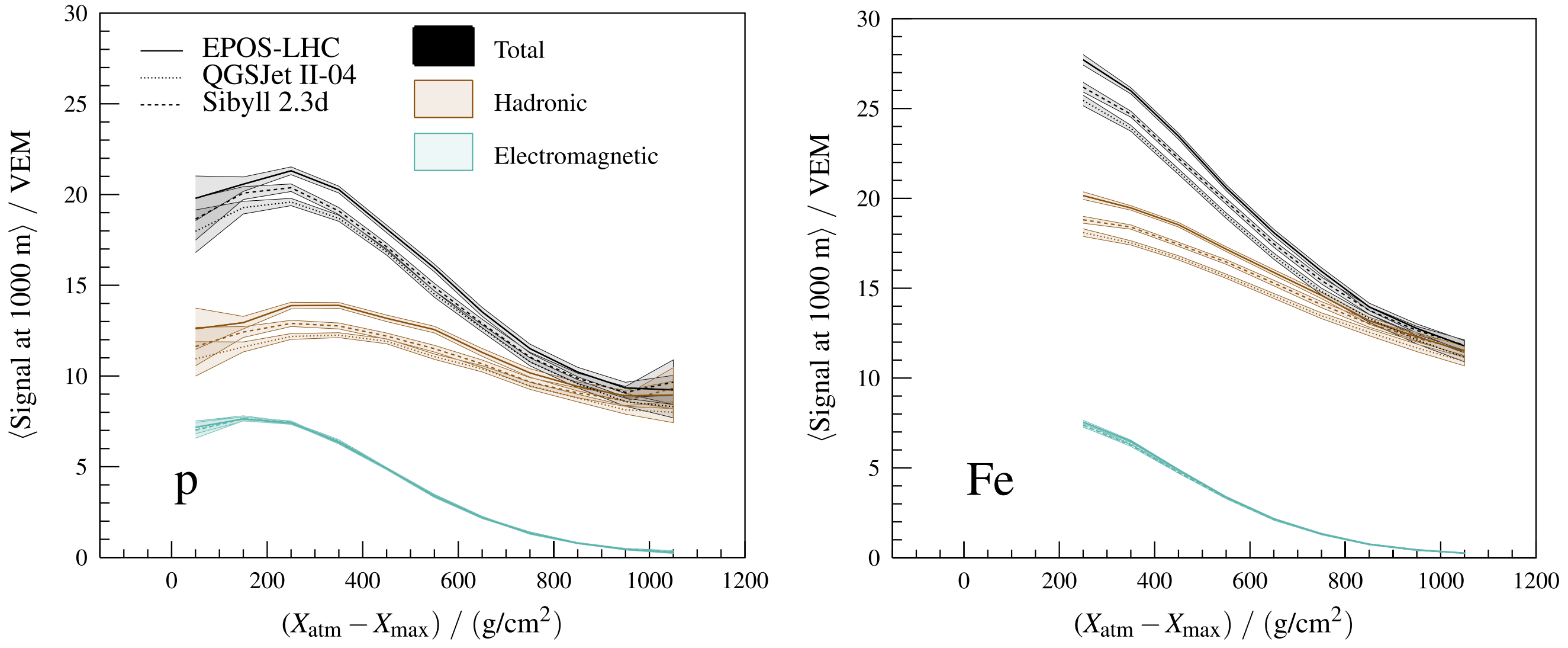

FIG. 2. The total ground signal at 1000 m from the shower core (black) and its hadronic (brown) and em (turquoise) components as a function of the distance from $X_{\max}$ to the ground in atmospheric depth units for protons (left) and iron nuclei (right) for different models. The bands contain the statistical uncertainty. $E = 10^{18.5}$ eV to $10^{19.0}$ eV, $\theta < 60°$.

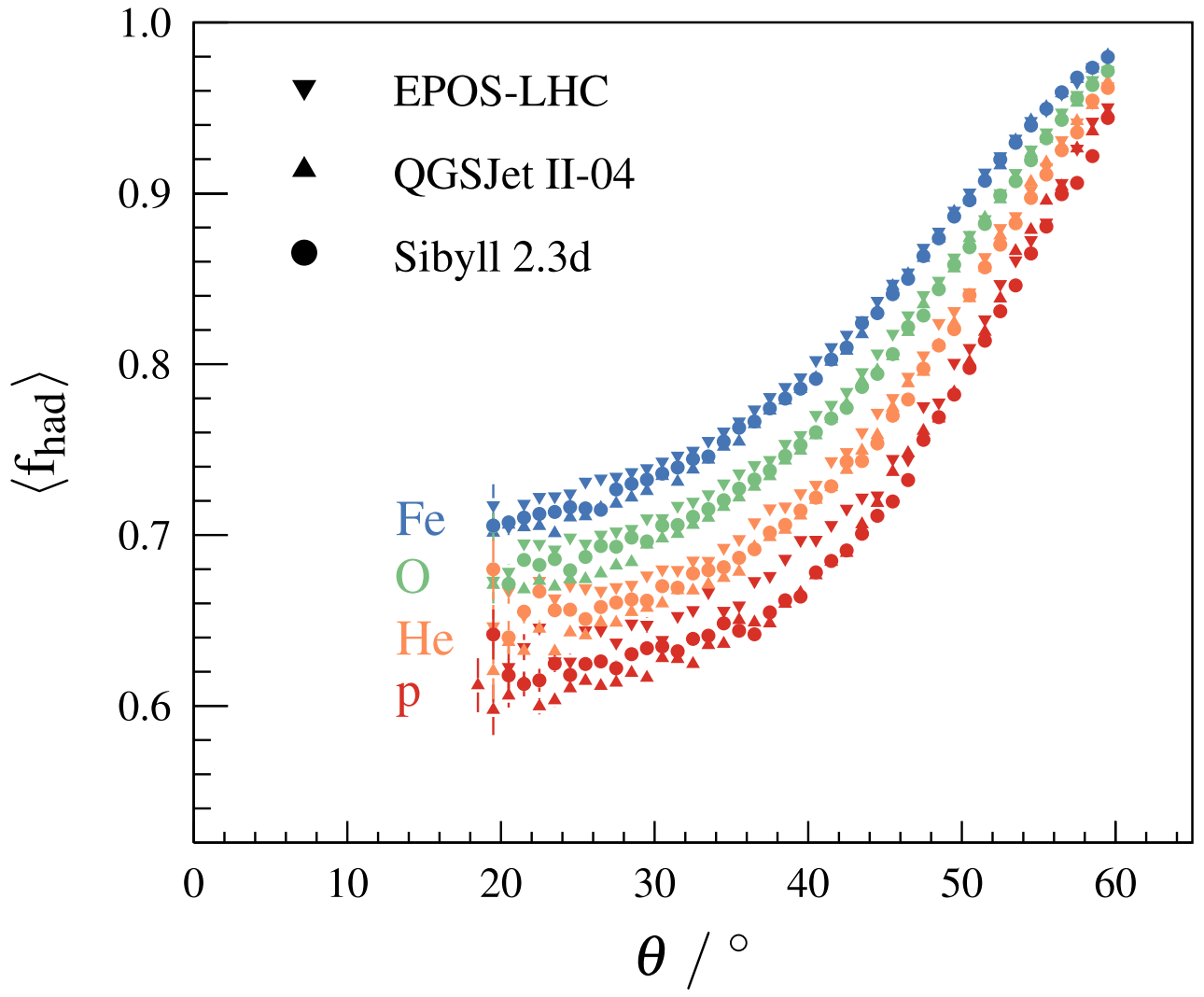

FIG. 3. The average fraction of hadronic signal at 1000 m from the shower core as a function of the reconstructed zenith angle for different models and primary masses. $E = 10^{18.5}$ eV to $10^{19.0}$ eV.

The average effect of the $X_{\max}$ change on the ground signal is incorporated through the separate effects on the em $g_{\mathrm{em},k}$ and hadronic $g_{\mathrm{had},k}$ signals, see Appendix B for details. We parametrized the evolution of the mean ground signal parts with the distance of $X_{\max}$ to the ground in atmospheric depth units, $X_{\mathrm{atm}} - X_{\max}$, where $X_{\mathrm{atm}} = 880\ \mathrm{g/cm^2}/\cos\theta$, see the examples in Fig. 2. In this way, the total ground signal is estimated to be modified (via $g_{\mathrm{em},k}$, $g_{\mathrm{had},k}$) at most by about 7% for a change of $X_{\max}$ by 50 g/cm$^2$. This is in accordance with the functional dependencies in Fig. 2 weighted over the relative contributions of hadronic and em signals (see Fig. 3) and over the primary fractions.

As a consequence of the four-component shower universality approach, the em signal is very similar in all three models, see Fig. 2. The differences in the total signal stem from the size of the hadronic signal at different zenith angles, corresponding to different $X_{\mathrm{atm}} - X_{\max}$ values. Therefore, the freedom in $R_{\mathrm{had}}(\theta)$ and also $\Delta X_{\max}$ (see Sec. I) removes the main differences in predictions of $S$ and $X$ for the three models. We assumed a linear dependence of $R_{\mathrm{had}}(\theta)$ on $X_{\mathrm{atm}} - X_{\max}$ and defined rescaling parameters of the hadronic signal at two extreme zenith angles, $R_{\mathrm{had}}(\theta_{\min})$ and $R_{\mathrm{had}}(\theta_{\max})$, for ~28° and ~55°, respectively (see Appendix B for the definition).

We have verified using MC-MC tests, see Appendix C, that the method is performing well and the observed biases were taken as sources of systematic uncertainty of the results.

[3]Note that the choice of $E^{\mathrm{ref}} \approx \langle E_{\mathrm{FD}} \rangle$ minimizes the effect of the energy factors; $(E^{\mathrm{ref}})^{\beta-1/B}/\langle E_{\mathrm{FD}}^{\beta-1/B} \rangle_k \approx 1$ and $(E^{\mathrm{ref}})^{1-1/B}/\langle E_{\mathrm{FD}}^{1-1/B} \rangle_k \approx 1$.

## III. DATA ANALYSIS

### A. Data selection

The analysis is applied to hybrid events, i.e., events detected with both SD and FD of the Pierre Auger Observatory between 1 January 2004 and 31 December 2018. To select high-quality events, for the FD part we apply the selection criteria used for the $X_{\max}$ analysis [9,11], for the SD part the selection is the same as in the measurements of the SD energy spectrum [22], additionally removing events with saturated SD stations. In this way, we ensure an accurate estimation of the observables $X_{\max}$, $E_{\mathrm{FD}}$, $\theta$, and $S(1000)$ used as the inputs to the method. The selection efficiency is similar for all four primary masses and all zenith-angle bins, therefore it does not introduce mass-composition biases. In total, 2239 hybrid events were selected in the FD energy range $10^{18.5}$ eV to $10^{19.0}$ eV ($\langle E_{\mathrm{FD}} \rangle \approx 10^{18.7}$ eV) and zenith angles between 0° and 60°. The data were divided into five zenith-angle bins containing nearly the same ($N = 425$–$500$) and sufficiently large number of events, see Fig. 4.

To take into account long-term performance of the FD and SD [34], we applied a time-dependent correction to $E_{\mathrm{FD}}$ with a negligible systematic effect on the final results. The signal $S(1000)$ was corrected for seasonal atmospheric effects [35].

### B. Results

The minimized values of the log-likelihood expression [see Eq. (3)] are summarized in Table II, progressively applying different modifications to the models. At all stages, the fits of the mass composition are performed with (p, He, O, Fe) fractions as the free-fit parameters. First, the model predictions without any modifications are used for fitting the data. Then we perform fits adding freedom in only one of the modifications $\Delta X_{\max}$, $R_{\mathrm{had}}$ (independent on zenith angle), $R_{\mathrm{had}}(\theta)$ (zenith-angle dependent). Finally, the fits are performed using combinations of ($R_{\mathrm{had}}$, $\Delta X_{\max}$) and ($R_{\mathrm{had}}(\theta)$, $\Delta X_{\max}$). From the minimum values of the log-likelihood expression for these scenarios, and from the likelihood-ratio test for the nested model, one can see that the major improvements in the data description are achieved due $R_{\mathrm{had}}$ and then $\Delta X_{\max}$ modifications.

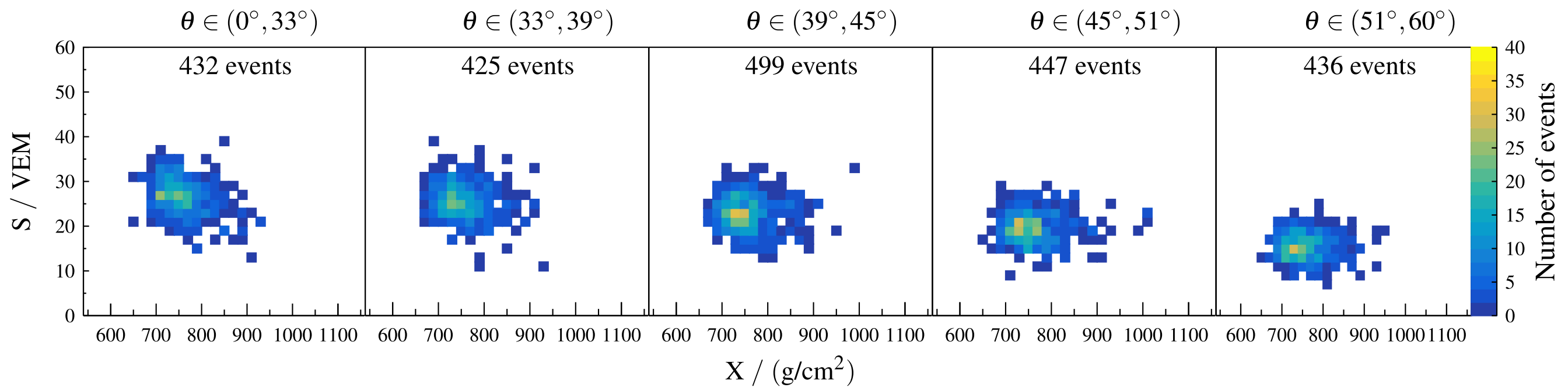

FIG. 4. Distributions of ground signal at 1000 m ($S$) and depth of shower maximum ($X$), see Eqs. (1) and (2), for the data of the Pierre Auger Observatory in the energy range $10^{18.5}$ eV to $10^{19.0}$ eV in five zenith-angle bins.

The improvement in the data description due to the introduction of the zenith-angle dependence in $R_{had}$ is statistically significant for QGSJet-II-04, and less significant for SIBYLL 2.3d, with a negligible effect in the case of Epos-LHC.

The measured two-dimensional ($S$, $X$) distributions are described acceptably by MC templates modified by $R_{had}(\theta)$ and $\Delta X_{max}$ in all five zenith-angle bins with $p$-values estimated to be higher than 10% for all three models. These probabilities are obtained from MC-MC tests (see Appendix C) using distributions of values of log-likelihood expression from the fitting of 500 random samples consisting of 2239 simulated showers. Each MC sample had a mass composition and artificially modified $X_{max}$ and $S_{had}(\theta)$ following the values obtained from the best fits to data. We plot in Fig. 5 projected distributions of $X$ and $S$ (together with plots in the logarithmic scale to stress the consistency also for the tails of distributions) and their zenith-angle dependent correlation using the Gideon-Hollister correlation coefficient [38] for the data and MC templates with the modifications of ($R_{had}(\theta)$, $\Delta X_{max}$). In this way, we demonstrate that with the modified MC templates, we achieve a consistent simultaneous description of $X_{max}$, $S(1000)$ and their correlation for the data shown in Fig. 4. The models without modifications or only with the zenith-angle independent modification of $R_{had}$ do not describe the data equivalently well (see Appendix D).

The resulting parameters of the data fits with ($R_{had}(\theta)$, $\Delta X_{max}$) modifications are presented in Table III. For all three models, a deeper $X_{max}$ scale is favored (see also left panels in Figs. 6 and 7) that would result in heavier primary mass composition derived from the $X_{max}$ data compared to the inferences with nonmodified predictions of the models [10,39]. The modifications $\Delta X_{max}$ are such that they reduce the difference in $X_{max}$ scales between the models (see Fig. 10 in Sec. IV) and, as a consequence, similar estimations of the fractions of the primary nuclei can be inferred from the data, see the right panel of Fig. 6.

Due to the shift of the $X_{max}$ predictions deeper in the atmosphere, more shower particles reach the ground producing a few percent larger SD signals compared to the nonmodified models. Therefore, the total increase of $S$ for the modified MC predictions consists of contributions from both $\Delta X_{max}$ (∼0 to 7%) and $R_{had}(\theta)$ (∼10 to 16%) as shown in the left panel of Fig. 8. The two effects from the modification of the $X_{max}$ scale—heavier primary mass composition and a larger number of shower particles reaching the ground—lead to the values of $R_{had}(\theta)$ (Table III) that are smaller than the values previously found for Epos-LHC and QGSJet-II-04 in Ref. [17]. The modified and original $S_{had}$ scales are shown for the three models in the right panel of Fig. 8.

TABLE II. Evolution of the minimum values of the log-likelihood expression [see Eq. (3)] fitting the data with different modifications of the model predictions. In all cases, except $R_{had}$ = const and $\Delta X_{max}$, the significance of improvement of data description with the $R_{had}(\theta)$ and $\Delta X_{max}$ fit is above $5\sigma$ using the likelihood-ratio test applying the Wilks' theorem [36] for nested model [37]. In the case of $R_{had}$ = const and $\Delta X_{max}$, the improvement of data description with $R_{had}(\theta)$ and $\Delta X_{max}$ fit is $\sim 0.1\sigma$, $\sim 4.4\sigma$, and $\sim 2.0\sigma$ for Epos-LHC, QGSJet-II-04, and SIBYLL 2.3d, respectively.

| $\ln \mathcal{L}_{min}$ | Epos-LHC | QGSJet-II-04 | SIBYLL 2.3d |
|---|---|---|---|
| None | 2022.9 | 4508.0 | 2496.5 |
| $\Delta X_{max}$ | 738.6 | 1674.8 | 1015.7 |
| $R_{had}$ = const | 489.2 | 684.4 | 521.6 |
| $R_{had}(\theta)$ | 489.2 | 673.9 | 517.6 |
| $R_{had}$ = const and $\Delta X_{max}$ | 452.2 | 486.7 | 454.2 |
| $R_{had}(\theta)$ and $\Delta X_{max}$ | 451.9 | 476.3 | 451.6 |

### C. Systematic uncertainties

There are four dominant sources of systematic uncertainties in the fitted parameters:

(i) The uncertainty in the FD energy scale ±14% [22];
(ii) The uncertainty in the $X_{max}$ measurements $^{+8}_{-9}$ g/cm$^2$ [9];
(iii) The uncertainty in the $S(1000)$ measurements ±5% [7];

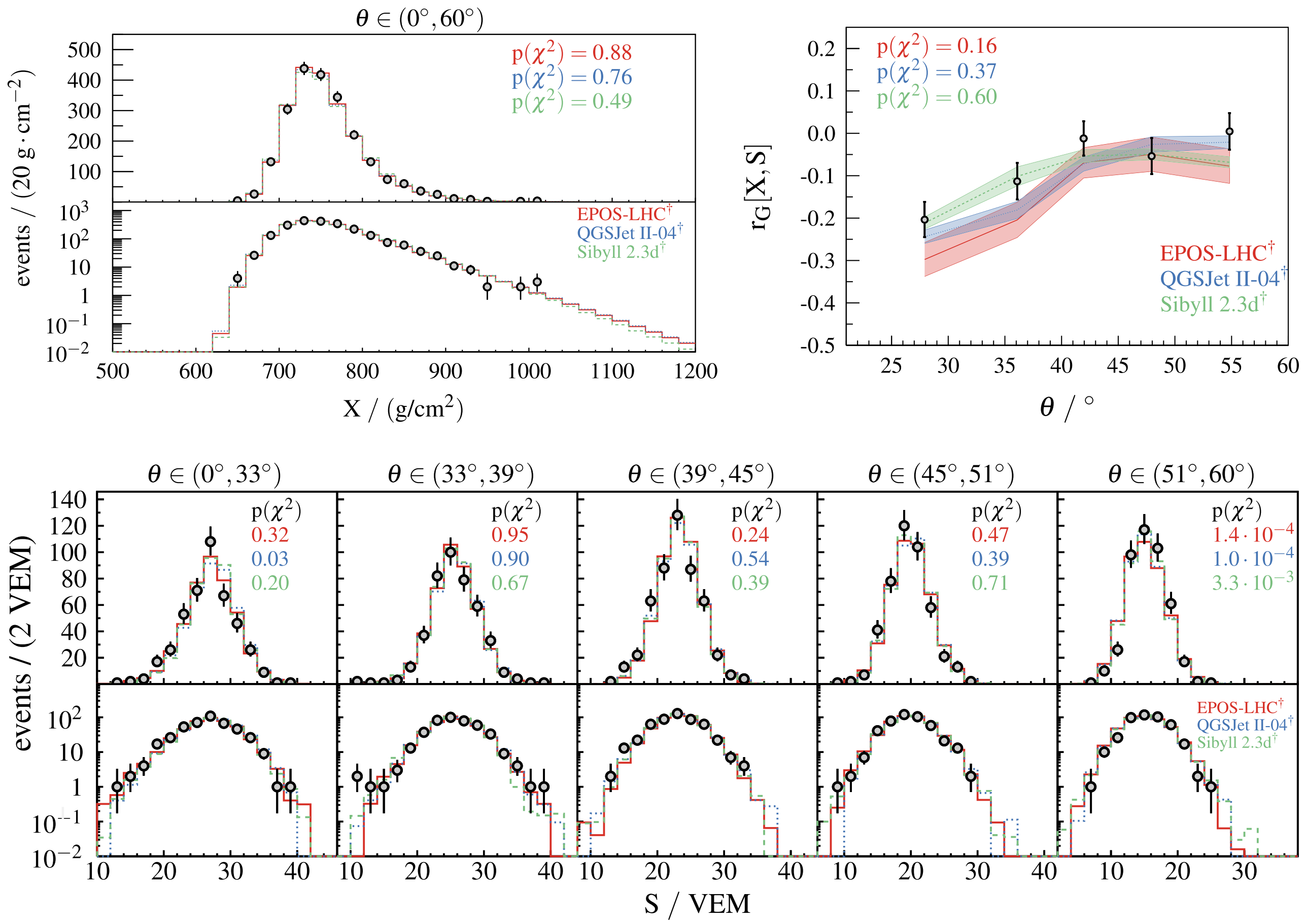

FIG. 5. Distributions of depth of shower maximum ($X$, top-left) and ground signal at 1000 m ($S$, bottom), see Eqs. (1) and (2), for the Auger data (points) and for the best fits of ($S$, $X$) distributions with $R_{\mathrm{had}}(\theta)$ and $\Delta X_{\mathrm{max}}$ modifications in predictions of three models (denoted by † as templates were modified from original predictions). In the top-right panel, the dependence of the Gideon-Hollister coefficient of correlation [38] between $S$ and $X$ on the zenith angle is shown. The $\chi^2$ probabilities characterize the compatibility between measurements and MC predictions in the individual plots.

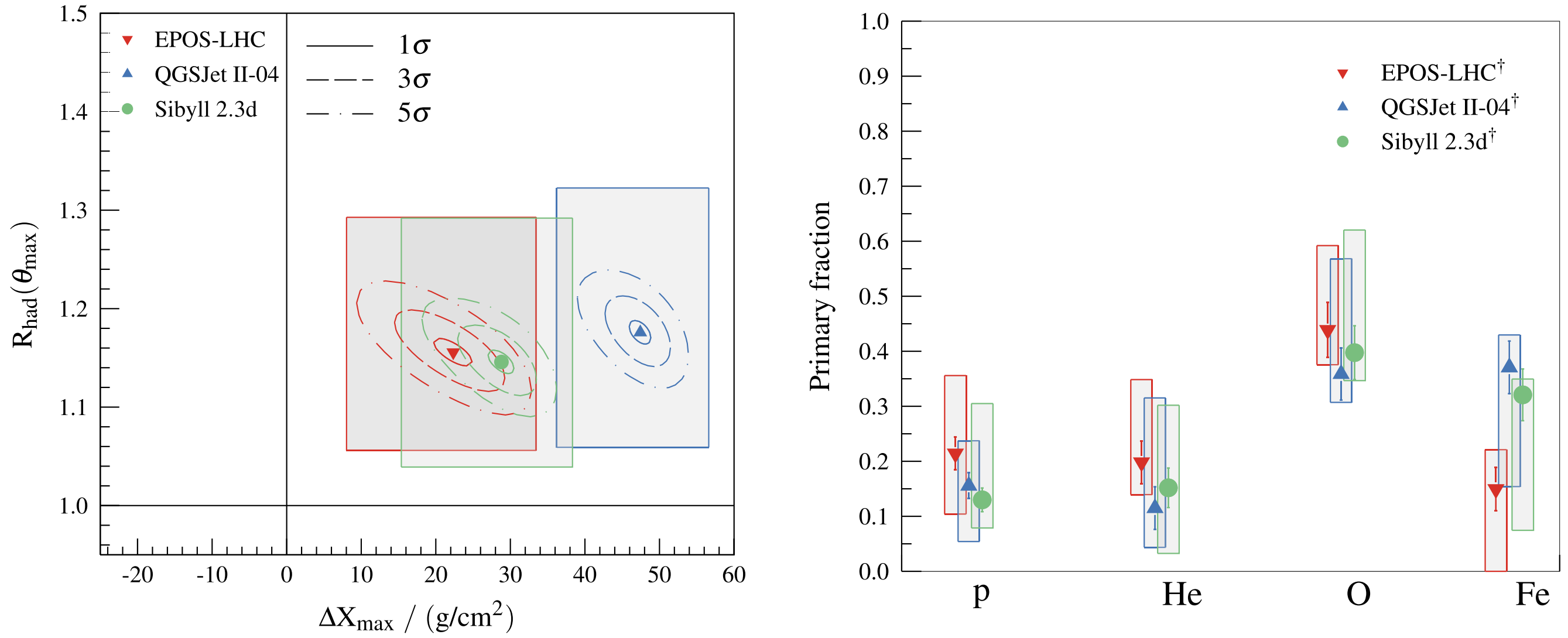

FIG. 6. *Left:* Correlations between $\Delta X_{\mathrm{max}}$ and $R_{\mathrm{had}}(\theta_{\mathrm{max}} \approx 55°)$ modifications of the model predictions obtained from the data fits. The contours correspond to $1\sigma$, $3\sigma$, and $5\sigma$ statistical uncertainties. The gray rectangles are the projections of the total systematic uncertainties. *Right:* The most likely primary fractions of the four components from the data fits using $\Delta X_{\mathrm{max}}$ and $R_{\mathrm{had}}(\theta)$. The height of the gray bands shows the size of projected total systematic uncertainties.

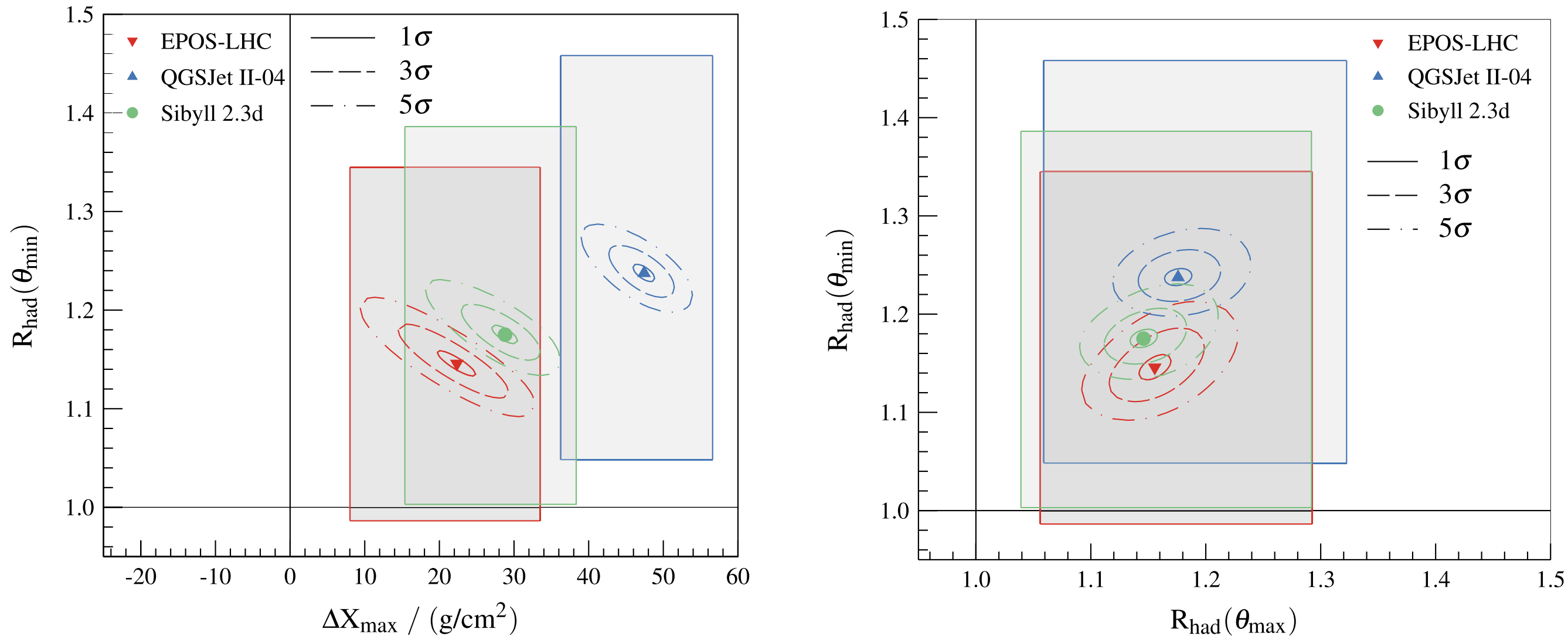

FIG. 7. *Left:* Correlations between $\Delta X_{max}$ and $R_{had}(\theta_{min} \approx 28°)$ modifications of the model predictions obtained from the data fits. *Right:* Correlation between $R_{had}(\theta_{max} \approx 55°)$ and $R_{had}(\theta_{min} \approx 28°)$. The contours correspond to $1\sigma$, $3\sigma$, and $5\sigma$ statistical uncertainties. The gray rectangles are the projections of the total systematic uncertainties.

(iv) The biases of the method estimated from the MC-MC tests (see Appendix B for the results of these tests).

Since the $X_{max}$ systematic uncertainty is strongly correlated with the modification $\Delta X_{max}$, its effect on the nuclei fractions is nearly cancelled out by the corresponding change of $\Delta X_{max}$. In general, the nuclei fractions, and therefore the inferences on the mass composition, are weakly sensitive to all experimental systematic errors due to the simultaneous fitting of $\Delta X_{max}$ and $R_{had}$ in the method. To explore the effect of those systematics, and as a simplifying ansatz, the total systematic uncertainties on the fit parameters are obtained by summing all four contributions in quadrature (see Appendix E for the size of individual contributions).

One can also note, that within systematic errors no significant dependence of $R_{had}$ on the zenith angle was found. The difference between $R_{had}(\theta_{min})$ and $R_{had}(\theta_{max})$ shows a tight correlation with the uncertainty on the energy scale. However, given all the experimental uncertainties in the case of QGSJet-II-04, the measured data prefers within the method rather flatter attenuation of the hadronic signal at 1000 m than predicted by the model, indicating too hard spectra of muons predicted by this model.

The systematic uncertainties on the parameters $B$, $D$, $\beta$, used in Eqs. (1) and (2) for the energy correction of $S$, $X$, as well as corrections of the long-term performance and other effects related to the operation of the SD and FD, have a negligible contribution to the systematic uncertainties. We could not identify any significant dependencies of the results on the zenith angle or energy in the studied ranges.

### D. Significance of improvement in data description

In Fig. 9, the results of our method for $\Delta X_{max}$ and $R_{had}(\theta)$ applying also all possible combinations of the systematic uncertainties on $E_{FD}$, $X_{max}$, and $S(1000)$ are shown with the full points. These points are located approximately in a plane, contour outlined with a dashed line, due to a correlation between the modification parameters through the mass composition describing the data (see the left panels of Figs. 6 and 7, e.g., increase of $\Delta X_{max}$ leads to a heavier fitted mass composition and consequently to a decrease of $R_{had}$). The plane is tilted with respect to the $[R_{had}(\theta_{min}), R_{had}(\theta_{max})]$ plane. It is a consequence of the effect of $\Delta X_{max}$ on the ground signal $S$ at different zenith angles, see Eqs. (B1)–(B5) in Appendix B, and consequently on the fitted $R_{had}(\theta_{min})$, $R_{had}(\theta_{max})$. The color of the points corresponds to the difference in fitted log-likelihood expressions ($\Delta \ln \mathcal{L}$) in case of no modifications and in case of the assumed template modifications. The closest approach to the point of no modification is estimated through a dense scan of linear combinations of experimental systematic uncertainties for the lowest values of $\Delta \ln \mathcal{L}$ that are, in some cases, even beyond the range of systematic uncertainties quoted by Auger (see Appendix F for more details). For all three models, the closest approach of $\Delta \ln \mathcal{L}$ (indicated by a line in Fig. 9 connecting point $[1, 1, 0\ \mathrm{g/cm^2}]$ with the plane) is $>19$ which is still higher than the value estimated using the Wilks' theorem in the likelihood-ratio test for nested model at the level of $5\sigma$ ($\Delta \ln \mathcal{L} \approx 16.62$). This confirms that the modifications of $X_{\mathrm{max}}$ and $R_{\mathrm{had}}$ scales are not an artifact of the systematic uncertainties of our measurements but a needed change in the model descriptions. A correction of the results for the biases seen in the MC-MC tests leads to even larger significance values.

TABLE III. Modifications of the model predictions and primary fractions in the energy range $10^{18.5}$ eV to $10^{19.0}$ eV with statistical and systematic uncertainties for the best data fits and the $p$-values obtained using MC-MC tests.

| | $R_{had}(\theta_{min})$ | $R_{had}(\theta_{max})$ | $\Delta X_{max}/(g/cm^2)$ | $f_p$ (%) | $f_{He}$ (%) | $f_O$ (%) | $f_{Fe}$ (%) | $p$-value (%) |
|---|---|---|---|---|---|---|---|---|
| Epos-LHC | $1.15 \pm 0.01^{+0.20}_{-0.16}$ | $1.16 \pm 0.01^{+0.14}_{-0.10}$ | $22 \pm 3^{+11}_{-14}$ | $21 \pm 3^{+14}_{-11}$ | $20 \pm 4^{+15}_{-6}$ | $44 \pm 5^{+15}_{-6}$ | $15 \pm 4^{+7}_{-15}$ | 10.6 |
| QGSJet-II-04 | $1.24 \pm 0.01^{+0.22}_{-0.19}$ | $1.18 \pm 0.01^{+0.15}_{-0.12}$ | $47^{+2}_{-1}{}^{+9}_{-11}$ | $16 \pm 2^{+8}_{-10}$ | $11 \pm 4^{+20}_{-7}$ | $36 \pm 5^{+21}_{-5}$ | $37 \pm 5^{+6}_{-22}$ | 19.8 |
| SIBYLL 2.3d | $1.18 \pm 0.01^{+0.21}_{-0.17}$ | $1.15 \pm 0.01^{+0.15}_{-0.11}$ | $29 \pm 2^{+10}_{-13}$ | $13 \pm 2^{+18}_{-5}$ | $15 \pm 4^{+15}_{-12}$ | $40 \pm 5^{+22}_{-5}$ | $32 \pm 5^{+3}_{-25}$ | 32.6 |

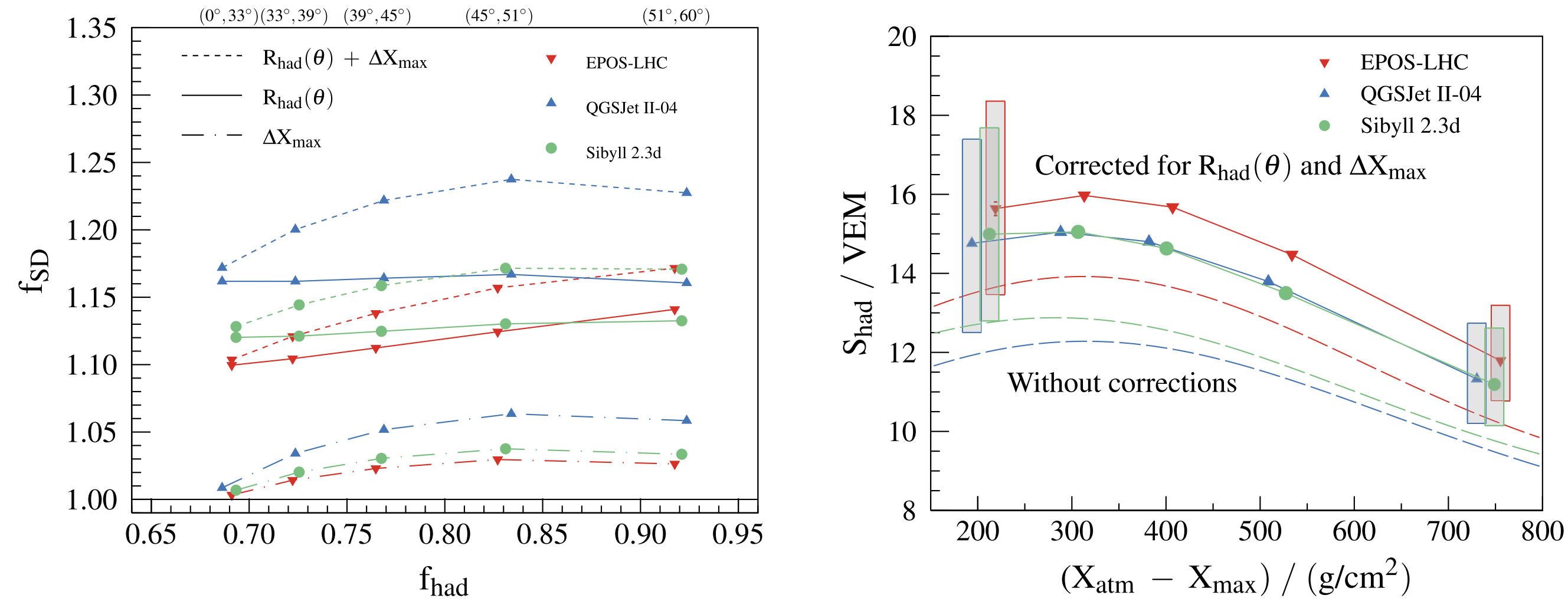

FIG. 8. *Left:* The total rescaling of $S$ (dashed lines) broken down into the contributions from the rescaling of the hadronic signal $R_{\mathrm{had}}(\theta)$ (solid lines) and the change of the predicted $X_{\mathrm{max}}$ scale ($\Delta X_{\mathrm{max}}$) (dash-dotted lines) for different models. *Right:* The dependence of the hadronic signal at 1000 m on the distance of $X_{\mathrm{max}}$ to the ground in atmospheric depth units as predicted for proton showers generated using nonmodified models (dashed lines) and accounting for the $R_{\mathrm{had}}(\theta)$, $\Delta X_{\mathrm{max}}$ modifications (solid lines). The height of the gray bands shows the size of projected total systematic uncertainties.

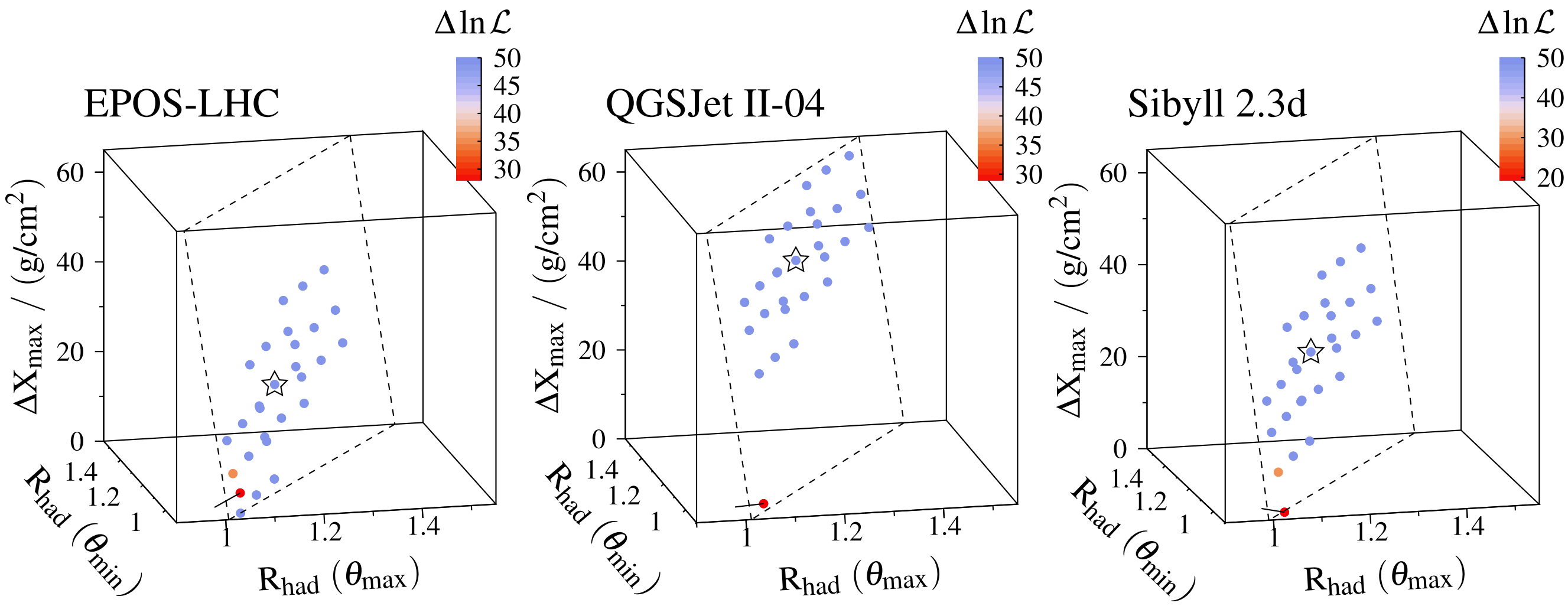

FIG. 9. Values of the modification parameters (points) for all possible combinations of experimental systematic uncertainties on the energy ($\pm14\%$), $X_{\mathrm{max}}$ ($^{+8}_{-9}$ g/cm$^2$), and $S(1000)$ ($\pm5\%$). The color of the points shows the difference in log-likelihood expressions ($\Delta \ln \mathcal{L}$) in the case of no modifications and in the case of the assumed template modifications, including the differences higher than 50 (note the slightly different scale between models). The results (see Table III) for no systematic shift of the data are highlighted by stars. Dashed lines outline the contour of the plane from the best-fit to the points. The closest approach to the nonmodified ($R_{\mathrm{had}}(\theta) = 1$, $\Delta X_{\mathrm{max}} = 0$ g/cm$^2$) model predictions using a dense scan of linear combinations of experimental systematic uncertainties is connected with this point by a black line. The animated rotated views are available at [40].

## IV. DISCUSSION

### A. Implications for inferences on mass composition

One straightforward consequence of the $X_{max}$ shift deeper in the atmosphere is the solution of the problem with the negative variance of the logarithmic mass $\sigma^2(\ln A)$ derived with QGSJet-II-04 from the measured $\langle X_{max}\rangle$ and $\sigma(X_{max})$ (as discussed in Sec. I). After application of the corresponding $\Delta X_{max}$ shifts, one finds $\sigma^2(\ln A) \approx 0.5$ to 2.5 in the energy range $10^{18.5}$ eV to $10^{19.0}$ eV for all models used in this work. These values are consistent with the degree of mixing of the primary composition found in the analysis of the correlation between $X_{max}$ and $S(1000)$ with nonmodified models, since the correlation analysis relies on the general phenomenology of air showers and this way is weakly sensitive to the uncertainties in the description of hadronic interactions [11,12].

Another outcome of the method can be foreseen by taking into account the quasiuniversal behavior of the $X_{max}$ elongation rate for all pure beams and models with values staying within 54 g/cm$^2$/decade to 61 g/cm$^2$/decade range. Changing the elongation rate within these limits introduces an energy-dependent uncertainty on the MC $X_{max}$ scale of about 4 g/cm$^2$/decade at most. Under the assumption that the difference $\Delta X_{max}$ between the models and data remains nearly independent of the primary energy, i.e., that there is no new physics that can significantly change the predictions for the $X_{max}$ elongation rate of single primary species, one could speculate that at the highest energies ($E \gtrsim 10^{19.5}$ eV) the Auger $X_{max}$ measurements, see Fig. 10, can be described with a heavy mass composition having a low degree of mixing due to $\langle X_{max}\rangle$ and $\sigma(X_{max})$ staying between the extrapolated predictions of the modified models for oxygen and iron nuclei.

### B. Primary species in the cosmic-ray beam

We checked if the shape of the data distribution of ground signal and $X_{max}$ and its zenith-angle evolution can be fitted better with an artificially reduced range of the masses in the MC templates and thus with different values of $R_{had}(\theta)$ and $\Delta X_{max}$. Due to the presence of deep events and $\sigma(X_{max})$ values close to the predictions for protons in $10^{18.5}$ eV to $10^{19.0}$ eV range, we kept protons in the MC templates and used (p, He, O) and (p, He) mixes for the data fits. In both cases, the quality of data fits was found to be inferior compared to the fits with (p, He, O, Fe) nuclei (see Table IV). This was confirmed by the MC-MC tests and was observed using the fits to the measured data for all three models that the obtained $X_{max}$ scale decreases by about 5 g/cm$^2$ to 7 g/cm$^2$, 10 g/cm$^2$ to 17 g/cm$^2$, and 30 g/cm$^2$ to 40 g/cm$^2$ and the hadronic signal scale by about 2% to 5%, 4% to 9%, and 15% to 20% when the heaviest primary Fe is replaced by Si, O, and He in the fit, respectively.

For the five-component (p, He, O, Si, Fe) fits, the values of $R_{had}(\theta)$ and $\Delta X_{max}$ remain well within statistical errors from the values obtained from the (p, He, O, Fe) fits. Though the fractions of silicon and iron nuclei are strongly anticorrelated in such fits, the silicon fraction remains low, <5%.

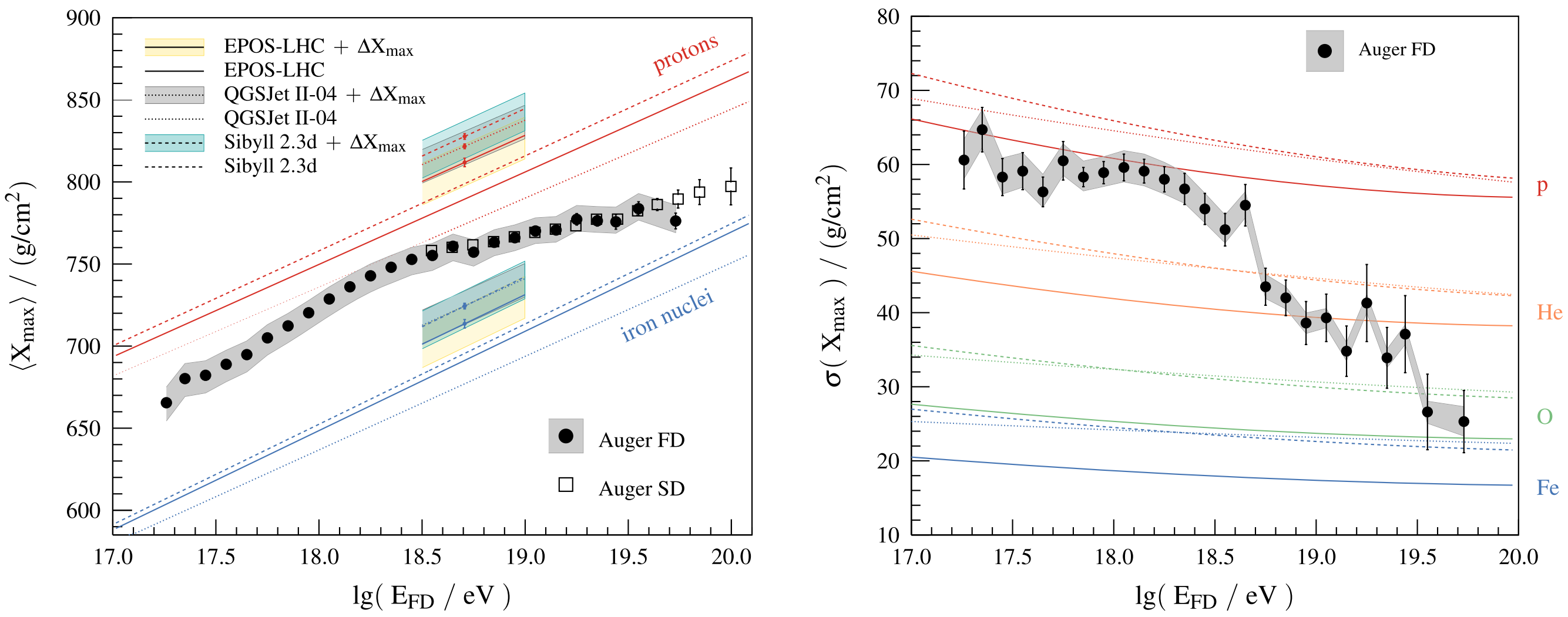

FIG. 10. The energy evolution of the mean (left) and the standard deviation (right) of the $X_{max}$ distributions measured by the Pierre Auger Observatory using FD [11] (solid circles) and SD [41] (open squares). The results of this paper for the modified $X_{max}$ scales (left panel) are shown with shaded bands with the heights corresponding to the systematic uncertainties. The original nonmodified model predictions for different primary species are shown with lines for the entire energy range.

TABLE IV. Minimum values of the log-likelihood expression [see Eq. (3)] for different number of primaries assumed to be present in the model predictions. In all cases, the significance of improvement of data description with the $R_{had}(\theta)$, $\Delta X_{max}$, and p, He, O, Fe fit using the likelihood-ratio test applying the Wilks' theorem for nested model is above $5\sigma$.

| $\ln \mathcal{L}_{min}$ | Epos-LHC | QGSJet-II-04 | SIBYLL 2.3d |
|---|---|---|---|
| p He | 518.3 | 633.5 | 563.5 |
| p He O | 467.5 | 523.3 | 486.6 |
| p He O Fe | 451.9 | 476.3 | 451.6 |

### C. Modifications of hadronic interactions

Assuming the same modifications of hadronic interaction features as in Ref. [42], the observed increase of the $X_{max}$ scale in MC predictions could be explained by an increase of the elasticity or a decrease of the multiplicity and cross section. A detailed study of the combinations of such modifications at the energy equivalent to this study is ongoing [43].

The elasticity is a very good candidate for a potential source of $X_{max}$ modification because there are no precise data to constrain this in models at high energy. It could be precisely measured only in the region where no detector at the LHC experiments exists. Consequently, the different models have relatively different predictions with large uncertainties.

A reduction of the multiplicity can also increase $X_{max}$ since the available energy is shared between fewer particles leading to higher $\pi^0$ energy at the first interaction. But at the same time, the same effect will reduce the energy available for muon production effectively increasing the tension in $R_{had}$.

The superposition model [44] makes an *ad hoc* modification of $p$-$p$ cross section that would explain a single change of $X_{max}$ scale for all primaries rather difficult. The change in the case of iron nuclei would correspond to a change of $p$-air cross section at the energy 56 times smaller, $\sim 9 \times 10^{16}$ eV, which starts to be in tension with the corresponding LHC measurements of $p$-$p$ collisions [45]. Remarkably, the most recent cross section measurements by the ALFA experiment are lower and more precise [46] than the one used to tune the models, possibly indicating an overestimation of the current cross section of the air-shower simulations.

In the case of the multiplicity and elasticity, we can consider the model of shower development from Ref. [23] also assuming the energy increase of the total multiplicity as $N \propto N_0 E^{\alpha}$ and a decrease of the elasticity with energy $\kappa \propto \kappa_0 E^{-\omega}$. We see that an *ad hoc* change of the normalization of multiplicity ($N_0$) or elasticity ($\kappa_0$) would modify the $X_{max}$ independently of the primary particle and energy,

$$X_{max}^{A} = X_1^A + X_0 \ln \frac{\kappa E}{A 2N \xi_c^{\pi}}$$
$$= X_1^A + X_0 \left[ (1 - \alpha - \omega) \ln \frac{E}{A \xi_c^{\pi}} + \ln \frac{\kappa_0}{N_0} \right], \quad (9)$$

where $\xi_c^{\pi}$ is the critical energy of pions at which their decay and interaction lengths are equal.

The hadronic signal can be increased by increasing the multiplicity or decreasing the ratio of neutral pions according to Ref. [42] but, as previously described, an increase in the multiplicity would decrease $X_{max}$ increasing the tension there. A change in the ratios of different hadrons, in particular with more strange particles, is a more likely possible explanation according to recent LHC data [47,48].

The energy spectrum of muons naturally influences the dependence of $S_{had}$ on $X_{atm} - X_{max}$. Though the systematic uncertainty of the energy scale limits a significant conclusion deduced from the method [see $R_{had}(\theta_{min}) - R_{had}(\theta_{max})$ in Appendix E], the QGSJet-II-04 model seems to predict too hard spectra of muons. For instance, a larger fraction of strange particles like kaons in the shower would lead to harder muon spectra because of the larger critical energy (smaller lifetime) of strange particles.

There are of course other possible modifications of hadronic interactions that could influence the observed differences between the predictions from the models and the measured data like cross sections of low-mass mesons, energy spectra of pions etc., (see e.g., Ref. [5]).

### D. Limitations of the method

There are remaining differences between the models and, correspondingly, with the data due to various limitations of the method. Though our approach leads to a reduction of the differences between models in $\langle X_{max} \rangle$, the obtained modifications $\Delta X_{max}$ do not cancel out these differences completely. In particular, the modified $X_{max}$ scale for Epos-LHC is shallower compared to QGSJet-II-04 and SIBYLL 2.3d (see the left panel of Fig. 10). We found, that a large part of this difference ($\sim 10$ g/cm$^2$) can be removed if an additional smearing of $X_{max}$ is applied to Epos-LHC showers to compensate for the smaller $X_{max}$ fluctuations predicted by this model in comparison to the other two (possibly partly due to strong defragmentation of nuclei in Epos-LHC [49], as recently confirmed by newer model EPOS-LHCR [50]). The remaining difference of about 5 g/cm$^2$ between the models might be due to the statistical errors on $\Delta X_{max}$ and differences between the models in the separation of the primary species in $\langle X_{max} \rangle$.

Possible dependencies of $\Delta X_{max}$, $R_{had}(\theta)$, and fluctuations of $X_{max}$ and $S(1000)$ on the primary mass are out of the scope of this paper. We checked that adding a linear mass dependence of $\Delta X_{max}$ into the method did not improve the fit significantly. The assumption of a mass-independent $\Delta X_{\max}$ used in this paper was mainly motivated by the similar differences in predicted $\langle X_{\max}\rangle$ for different primaries [50]. Such a quasiuniversal difference is a consequence of the very similar energy dependencies of interaction features like multiplicity, elasticity, and cross section assumed in the models [51]. It means that an offset in these features would lead to an approximately mass-independent difference in the $X_{\max}$ predictions. An *ad hoc* modification of the energy dependence of these features, like in [42,43], would lead to mass-dependent $X_{\max}$ shift.

## V. CONCLUSIONS

In this paper, we tested the predictions of the models QGSJet-II-04, SIBYLL 2.3d, Epos-LHC regarding the depths of maximum of air-shower profiles $X_{\max}$ and the signal produced by air-shower particles in water-Cherenkov stations at 1000 m from the shower core, $S(1000)$, composed of electromagnetic and hadronic ($S_{\text{had}}$) parts. The test consisted of fitting with MC templates of two-dimensional distributions of ($S(1000)$, $X_{\max}$) measured at the Pierre Auger Observatory and obtaining the scales of $X_{\max}$ and $S_{\text{had}}(\theta)$ predicted by the models, as well as the fractions of four primary nuclei (p, He, O, Fe).

We found that for the best description of the data distributions in the energy range $10^{18.5}$ eV to $10^{19.0}$ eV for $\theta < 60°$ the MC predictions of $X_{\max}$ should be deeper in the atmosphere by about 20 $\text{g/cm}^2$ to 50 $\text{g/cm}^2$, and the hadronic signal should be increased by about 15% to 25%. These modifications reduce the differences between the models in $X_{\max}$ and $S(1000)$, and as a consequence, lead to smaller uncertainties on the estimated fractions of the primary nuclei. Due to the deeper MC $X_{\max}$ scale and, correspondingly, a heavier mass composition inferred from the data compared with nonmodified models, the scaling factors for the hadronic signal are found to be smaller than in previous estimations not considering any modifications to the MC $X_{\max}$ scales. The statistical significance of the improvement in the data description using the assumed modifications to the MC templates is above $5\sigma$ for all three models even accounting for all possible linear combinations of experimental systematic uncertainties. The difference in $R_{\text{had}}(\theta)$ at the two extreme zenith angles implies an indication that softer spectra of muons generated by QGSJet-II-04 in 1000 m from the shower core would describe the Auger data better.

The specific ways to produce the required changes in the models might consist in combinations of modifications of integral (cross section, multiplicity, elasticity, etc.) or differential (secondary particles energy spectra) characteristics of the hadronic interactions as discussed e.g., in Refs. [5,42,43]. Our method addresses only the first-order differences in the mean values of $X_{\max}$ and hadronic signal without taking into account their possible dependencies on the primary mass or energy. These dependencies, as well as the investigation of a modification of fluctuations of air-shower observables, need to be studied further to corroborate the scale adjustments found in this paper, supplemented also by future data with increased mass-sensitivity from AugerPrime [52], the upgrade of the Pierre Auger Observatory.

## ACKNOWLEDGMENTS

The successful installation, commissioning, and operation of the Pierre Auger Observatory would not have been possible without the strong commitment and effort from the technical and administrative staff in Malargüe. We are very grateful to the following agencies and organizations for financial support. Argentina—Comisión Nacional de Energía Atómica; Agencia Nacional de Promoción Científica y Tecnológica (ANPCyT); Consejo Nacional de Investigaciones Científicas y Técnicas (CONICET); Gobierno de la Provincia de Mendoza; Municipalidad de Malargüe; NDM Holdings and Valle Las Leñas; in gratitude for their continuing cooperation over land access; Australia—the Australian Research Council; Belgium—Fonds de la Recherche Scientifique (FNRS); Research Foundation Flanders (FWO), Marie Curie Action of the European Union Grant No. 101107047; Brazil—Conselho Nacional de Desenvolvimento Científico e Tecnológico (CNPq); Financiadora de Estudos e Projetos (FINEP); Fundação de Amparo à Pesquisa do Estado de Rio de Janeiro (FAPERJ); São Paulo Research Foundation (FAPESP) Grants No. 2019/10151-2, No. 2010/07359-6, and No. 1999/05404-3; Ministério da Ciência, Tecnologia, Inovações e Comunicações (MCTIC); Czech Republic—No. GACR 24-13049S, No. CAS LQ100102401, No. MEYS LM2023032, No. CZ.02.1.01/0.0/0.0/16_013/0001402, No. CZ.02.1.01/0.0/0.0/18_046/0016010, No. CZ.02.1.01/0.0/0.0/17_049/0008422 and No. CZ.02.01.01/00/22_008/0004632; France—Centre de Calcul IN2P3/CNRS; Centre National de la Recherche Scientifique (CNRS); Conseil Régional Ile-de-France; Département Physique Nucléaire et Corpusculaire (PNC-IN2P3/CNRS); Département Sciences de l'Univers (SDU-INSU/CNRS); Institut Lagrange de Paris (ILP) Grant No. LABEX ANR-10-LABX-63 within the Investissements d'Avenir Programme Grant No. ANR-11-IDEX-0004-02; Germany—Bundesministerium für Bildung und Forschung (BMBF); Deutsche Forschungsgemeinschaft (DFG); Finanzministerium Baden-Württemberg; Helmholtz Alliance for Astroparticle Physics (HAP); Helmholtz-Gemeinschaft Deutscher Forschungszentren (HGF); Ministerium für Kultur und Wissenschaft des Landes Nordrhein-Westfalen; Ministerium für Wissenschaft, Forschung und Kunst des Landes Baden-Württemberg; Italy—Istituto Nazionale di Fisica Nucleare (INFN); Istituto Nazionale di Astrofisica (INAF); Ministero dell'Università e della Ricerca (MUR); CETEMPS Center of Excellence; Ministero degli Affari Esteri (MAE), ICSC

Centro Nazionale di Ricerca in High Performance Computing, Big Data and Quantum Computing, funded by European Union NextGenerationEU, reference code CN_00000013; México—Consejo Nacional de Ciencia y Tecnología (CONACYT) No. 167733; Universidad Nacional Autónoma de México (UNAM); PAPIIT DGAPA-UNAM; The Netherlands—Ministry of Education, Culture and Science; Netherlands Organisation for Scientific Research (NWO); Dutch national e-infrastructure with the support of SURF Cooperative; Poland—Ministry of Education and Science, grants No. DIR/WK/2018/11 and 2022/WK/12; National Science Centre, Grants No. 2016/22/M/ST9/00198, No. 2016/23/B/ST9/01635, No. 2020/39/B/ST9/01398, and No. 2022/45/B/ST9/02163; Portugal—Portuguese national funds and FEDER funds within Programa Operacional Factores de Competitividade through Fundação para a Ciência e a Tecnologia (COMPETE); Romania—Ministry of Research, Innovation and Digitization, CNCS-UEFISCDI, Contract No. 30N/2023 under Romanian National Core Program LAPLAS VII, Grant No. PN 23 21 01 02 and Project No. PN-III-P1-1.1-TE-2021-0924/TE57/2022, within PNCDI III; Slovenia—Slovenian Research Agency, Grants No. P1-0031, No. P1-0385, No. I0-0033, No. N1-0111; Spain—Ministerio de Economía, Industria y Competitividad (FPA2017-85114-P and PID2019–104676 GB-C32), Xunta de Galicia (ED431C 2017/07), Junta de Andalucía (SOMM17/6104/UGR, P18-FR-4314) Feder Funds, RENATA Red Nacional Temática de Astropartículas (FPA2015-68783-REDT) and María de Maeztu Unit of Excellence (MDM-2016-0692); USA—Department of Energy, Contracts No. DE-AC02-07CH11359, No. DE-FR02-04ER41300, No. DE-FG02-99ER41107, and No. DE-SC0011689; National Science Foundation, Grant No. 0450696; The Grainger Foundation; Marie Curie-IRSES/EPLANET; European Particle Physics Latin American Network; and UNESCO.

## Appendix A: PARAMETRIZATION OF MC TEMPLATES

The MC templates of the two-dimensional distributions of $(S,\, X)$ normalized by the total number of simulated showers $N_{\mathrm{MC}}$ and weighted to correspond to the measured energy spectrum are fitted using the following function:

$$\Phi = \frac{\mathrm{d}N}{N_{\mathrm{MC}}\mathrm{d}X\mathrm{d}S} = A_{\mathrm{Gauss}} A_{\mathrm{Gumbel}} f_{\mathrm{Gumbel}}(X) f_{\mathrm{Gauss}}(X, S). \quad \text{(A1)}$$

The $X_{\mathrm{max}}$ part is described by the generalized Gumbel distribution [53],

$$f_{\mathrm{Gumbel}}(X) = \exp[-\lambda(x - \exp x))], \quad \text{(A2)}$$

where $x = (X - m)/s$ and the ground-signal part is assumed to follow the Gaussian distribution with the mean value linearly dependent on $X$

TABLE V. Parameters of MC templates for air showers generated with model Epos-LHC and initiated by a primary particle $i$, see Eq. (A1).

| $i$ | $\theta$ | $f_{\mathrm{had}}$ | $m/(\mathrm{g/cm^2})$ | $s/(\mathrm{g/cm^2})$ | $\lambda$ | $p/(\mathrm{VEM}/(\mathrm{g/cm^2}))$ | $q/\mathrm{VEM}$ | $r/\mathrm{VEM}$ |
|---|---|---|---|---|---|---|---|---|
| p | (0°, 33°) | 0.64 | 757 | 49 | 1.2 | $-1.6\times10^{-2}$ | 3.10 | 34.3 |
| p | (33°, 39°) | 0.67 | 755 | 44 | 1.1 | $-3\times10^{-3}$ | 2.85 | 22.9 |
| p | (39°, 45°) | 0.71 | 755 | 43 | $8.9\times10^{-1}$ | $6.7\times10^{-3}$ | 2.88 | 13.8 |
| p | (45°, 51°) | 0.77 | 754 | 47 | $9.8\times10^{-1}$ | $1.2\times10^{-2}$ | 2.83 | 6.3 |
| p | (51°, 60°) | 0.88 | 757 | 53 | 1.3 | $1.1\times10^{-2}$ | 2.68 | 3.0 |
| He | (0°, 33°) | 0.67 | 741 | 51 | 2.1 | $-2.7\times10^{-3}$ | 2.78 | 25.3 |
| He | (33°, 39°) | 0.70 | 739 | 49 | 2.0 | $7.7\times10^{-3}$ | 2.60 | 16.1 |
| He | (39°, 45°) | 0.75 | 742 | 61 | 2.6 | $1.7\times10^{-2}$ | 2.47 | 7.2 |
| He | (45°, 51°) | 0.81 | 741 | 57 | 2.5 | $1.7\times10^{-2}$ | 2.55 | 3.5 |
| He | (51°, 60°) | 0.91 | 739 | 64 | 2.7 | $1.3\times10^{-2}$ | 2.50 | 2.6 |
| O | (0°, 33°) | 0.71 | 714 | 63 | 5.7 | $-1.6\times10^{-3}$ | 2.61 | 26.2 |
| O | (33°, 39°) | 0.74 | 714 | 65 | 5.4 | $8.8\times10^{-3}$ | 2.45 | 16.9 |
| O | (39°, 45°) | 0.78 | 714 | 66 | 5.5 | $1.9\times10^{-2}$ | 2.41 | 6.8 |
| O | (45°, 51°) | 0.84 | 711 | 48 | 2.9 | $2.2\times10^{-2}$ | 2.37 | 1.8 |
| O | (51°, 60°) | 0.93 | 714 | 82 | 7.5 | $1.7\times10^{-2}$ | 2.33 | 1.0 |
| Fe | (0°, 33°) | 0.74 | 688 | 76 | 13.2 | $1.4\times10^{-2}$ | 2.65 | 17.4 |
| Fe | (33°, 39°) | 0.77 | 688 | 100 | 19.4 | $1.3\times10^{-2}$ | 2.62 | 15.7 |
| Fe | (39°, 45°) | 0.81 | 687 | 81 | 13.4 | $1.4\times10^{-2}$ | 2.51 | 11.9 |
| Fe | (45°, 51°) | 0.87 | 688 | 134 | 32.7 | $1.8\times10^{-2}$ | 2.41 | 6.2 |
| Fe | (51°, 60°) | 0.94 | 687 | 176 | 48.1 | $1.3\times10^{-2}$ | 2.34 | 4.9 |

TABLE VI. Same as in Table V, but for model QGSJet-II-04.

| $i$ | $\theta$ | $f_{\mathrm{had}}$ | $m/(\mathrm{g/cm^2})$ | $s/(\mathrm{g/cm^2})$ | $\lambda$ | $p/(\mathrm{VEM}/(\mathrm{g/cm^2}))$ | $q/\mathrm{VEM}$ | $r/\mathrm{VEM}$ |
|---|---|---|---|---|---|---|---|---|
| p | (0°, 33°) | 0.62 | 741 | 49 | 1.1 | $-1\times10^{-2}$ | 2.67 | 27.6 |
| p | (33°, 39°) | 0.65 | 748 | 70 | 1.9 | $1.1\times10^{-3}$ | 2.53 | 18.0 |
| p | (39°, 45°) | 0.69 | 743 | 56 | 1.3 | $1.2\times10^{-2}$ | 2.54 | 8.0 |
| p | (45°, 51°) | 0.76 | 739 | 47 | $9.5\times10^{-1}$ | $1.6\times10^{-2}$ | 2.51 | 2.2 |
| p | (51°, 60°) | 0.88 | 737 | 50 | 1.0 | $1\times10^{-2}$ | 2.35 | 2.7 |
| He | (0°, 33°) | 0.65 | 727 | 68 | 2.7 | $3.3\times10^{-3}$ | 2.49 | 19.0 |
| He | (33°, 39°) | 0.69 | 729 | 88 | 4.7 | $1.2\times10^{-2}$ | 2.35 | 11.0 |
| He | (39°, 45°) | 0.74 | 727 | 74 | 3.1 | $1.8\times10^{-2}$ | 2.38 | 4.5 |
| He | (45°, 51°) | 0.80 | 725 | 60 | 2.3 | $2\times10^{-2}$ | 2.20 | $4.6\times10^{-1}$ |
| He | (51°, 60°) | 0.91 | 724 | 74 | 3.0 | $1.1\times10^{-2}$ | 2.21 | 3.2 |
| O | (0°, 33°) | 0.69 | 700 | 99 | 9.5 | $1.1\times10^{-2}$ | 2.40 | 15.1 |
| O | (33°, 39°) | 0.73 | 702 | 87 | 6.6 | $1.5\times10^{-2}$ | 2.38 | 10.9 |
| O | (39°, 45°) | 0.77 | 702 | 91 | 7.5 | $2.1\times10^{-2}$ | 2.27 | 4.0 |
| O | (45°, 51°) | 0.84 | 704 | 102 | 8.0 | $2.1\times10^{-2}$ | 2.22 | $8.4\times10^{-1}$ |
| O | (51°, 60°) | 0.93 | 701 | 84 | 5.6 | $1.5\times10^{-2}$ | 2.15 | 1.8 |
| Fe | (0°, 33°) | 0.72 | 675 | 68 | 6.7 | $1.6\times10^{-2}$ | 2.41 | 13.9 |
| Fe | (33°, 39°) | 0.76 | 676 | 77 | 8.2 | $2.1\times10^{-2}$ | 2.37 | 8.2 |
| Fe | (39°, 45°) | 0.81 | 676 | 91 | 11.9 | $2.2\times10^{-2}$ | 2.37 | 4.7 |
| Fe | (45°, 51°) | 0.87 | 675 | 87 | 10.7 | $1.6\times10^{-2}$ | 2.24 | 6.4 |
| Fe | (51°, 60°) | 0.94 | 675 | 162 | 31.6 | $1.1\times10^{-2}$ | 2.19 | 5.6 |

$$f_{\mathrm{Gauss}}(X,S)=\exp\left[-\frac{y^2}{2r^2}\right], \tag{A3}$$

where $y=S-pX-q$. The normalization terms are given by

$$A_{\mathrm{Gauss}}=\frac{1}{\sqrt{2\pi}r} \quad \text{and} \quad A_{\mathrm{Gumbel}}=\frac{\lambda^\lambda}{s\Gamma(\lambda)}, \tag{A4}$$

where $\Gamma$ is the gamma function. The six free parameters in each of the MC template fits are $m$, $s$, $\lambda$ of the generalized

TABLE VII. Same as in Table V, but for model SIBYLL 2.3d.

| $i$ | $\theta$ | $f_{\mathrm{had}}$ | $m/(\mathrm{g/cm^2})$ | $s/(\mathrm{g/cm^2})$ | $\lambda$ | $p/(\mathrm{VEM}/(\mathrm{g/cm^2}))$ | $q/\mathrm{VEM}$ | $r/\mathrm{VEM}$ |
|---|---|---|---|---|---|---|---|---|
| p | 0°–33° | 0.63 | 769 | 70 | 2.1 | $-9\times10^{-3}$ | 2.97 | 27.7 |
| p | 33°–39° | 0.65 | 771 | 68 | 1.8 | $7.8\times10^{-4}$ | 2.96 | 19.0 |
| p | 39°–45° | 0.69 | 772 | 71 | 2.0 | $1.1\times10^{-2}$ | 2.84 | 9.0 |
| p | 45°–51° | 0.76 | 766 | 62 | 1.5 | $1.4\times10^{-2}$ | 2.73 | 4.4 |
| p | 51°–60° | 0.87 | 766 | 52 | 1.1 | $9.2\times10^{-3}$ | 2.77 | 4.0 |
| He | 0°–33° | 0.66 | 747 | 56 | 2.2 | $-1.3\times10^{-3}$ | 2.68 | 23.3 |
| He | 33°–39° | 0.69 | 745 | 50 | 1.8 | $1\times10^{-2}$ | 2.55 | 13.1 |
| He | 39°–45° | 0.73 | 742 | 54 | 1.8 | $1.6\times10^{-2}$ | 2.51 | 6.8 |
| He | 45°–51° | 0.80 | 744 | 58 | 2.1 | $2\times10^{-2}$ | 2.41 | $7.7\times10^{-1}$ |
| He | 51°–60° | 0.90 | 743 | 58 | 2.1 | $1.5\times10^{-2}$ | 2.41 | $6\times10^{-1}$ |
| O | 0°–33° | 0.70 | 715 | 48 2.3 | $8.2\times10^{-3}$ | 2.44 | 18.1 | |
| O | 33°–39° | 0.73 | 716 | 65 | 4.3 | $1.6\times10^{-2}$ | 2.41 | 10.8 |
| O | 39°–45° | 0.77 | 716 | 62 | 3.6 | $2.1\times10^{-2}$ | 2.36 | 4.9 |
| O | 45°–51° | 0.84 | 717 | 57 | 3.0 | $1.8\times10^{-2}$ | 2.26 | 3.8 |
| O | 51°–60° | 0.93 | 717 | 67 | 3.9 | $1.5\times10^{-2}$ | 2.25 | 1.8 |
| Fe | 0°–33° | 0.73 | 688 | 60 | 5.6 | $1.4\times10^{-2}$ | 2.61 | 16.2 |
| Fe | 33°–39° | 0.77 | 690 | 62 | 5.8 | $2.2\times10^{-2}$ | 2.40 | 8.6 |
| Fe | 39°–45° | 0.81 | 690 | 60 | 5.5 | $2.3\times10^{-2}$ | 2.40 | 5.5 |
| Fe | 45°–51° | 0.87 | 690 | 97 | 12.1 | $2\times10^{-2}$ | 2.35 | 3.8 |
| Fe | 51°–60° | 0.94 | 690 | 96 | 11.8 | $1.9\times10^{-2}$ | 2.24 | $8.7\times10^{-1}$ |

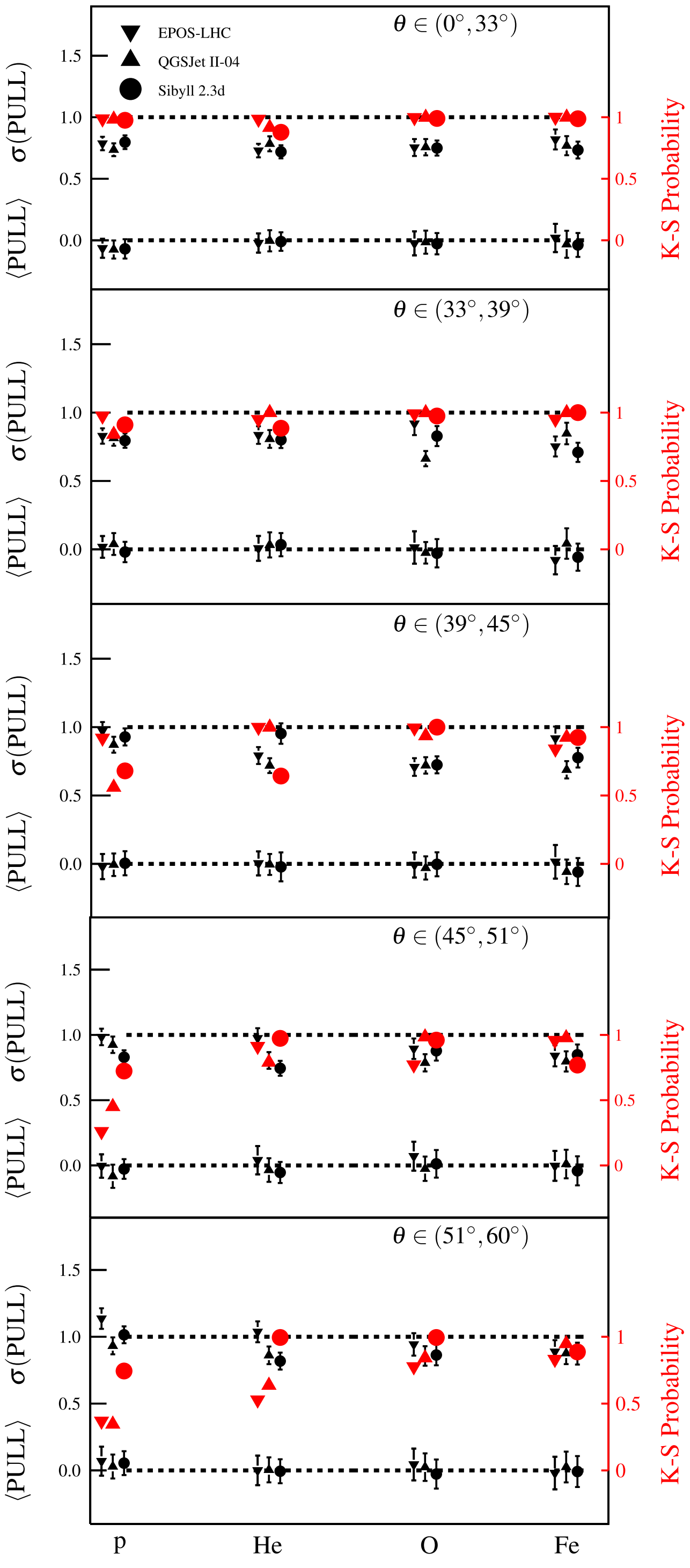

FIG. 11. The mean and the standard deviation of the MC distribution of $\mathrm{PULL} = (N_{\mathrm{fit}} - N_{\mathrm{mc}})/\sqrt{N_{\mathrm{mc}}}$ where $N_{\mathrm{fit}}$ is the value of the fitted function Eq. (A1) and $N_{\mathrm{mc}}$ is the value for each bin of the two-dimensional MC distribution $(S, X)$. The probability of consistency between the parametrized function and MC distribution (red) is tested with the two-dimensional Kolmogorov-Smirnov (K-S) test. $E = 10^{18.5}$ eV to $10^{19.0}$ eV.

Gumbel distribution, and $p$, $q$, $r$ of the Gaussian part. These fitted parameters for the three models used in this work are listed in Tables V–VII.

In Fig. 11, we show the mean and the standard deviation of the pull distribution for the description of each bin of the two-dimensional distribution $(S, X)$, see examples in Fig. 1, by the fits with a function Eq. (A1). The goodness of the description of the MC templates with these fits is also tested with the two-dimensional Kolmogorov-Smirnov test [54] demonstrating very good consistency between the MC templates and MC distributions of $(S, X)$.

## Appendix B: PARAMETRIZATION OF ATTENUATION OF GROUND SIGNALS

The electromagnetic and hadronic signals at 1000 m were corrected for the energy evolution as $S_{\mathrm{em}}^{\mathrm{ref}} = S_{\mathrm{em}}(E^{\mathrm{ref}}/E_{\mathrm{FD}})$ and $S_{\mathrm{had}}^{\mathrm{ref}} = S_{\mathrm{had}}(E^{\mathrm{ref}}/E_{\mathrm{FD}})^{1/\beta}$. The dependence of these average signals on the distance of $X_{\mathrm{max}}$ to the ground in atmospheric depth units, $t = X_{\mathrm{atm}}(\theta) - X_{\mathrm{max}}$, see Fig. 2, was parametrized with the Gaisser-Hillas function [55] allowing its vertical offset,

$$\langle S_{\alpha}^{\mathrm{ref}} \rangle(t) = S_{\alpha}^{0} \left( \frac{t - t_{\alpha}}{u_{\alpha} - t_{\alpha}} \right)^{Z(t_{\alpha})} \exp(Z(t)) + w_{\alpha}, \qquad \text{(B1)}$$

where $\alpha = \mathrm{had}$ or em, and normalization scaling $Z(x) = (u_{\alpha} - x)/v_{\alpha}$. $t_{\alpha}$ is the value of $X_{\mathrm{atm}}(\theta) - X_{\mathrm{max}}$ where the function reaches its maximum, $u_{\alpha}$ and $v_{\alpha}$ are parameters without a straightforward physics interpretation, and $S_{\alpha}^{0}$, $w_{\alpha}$ are the rescale and offset parameters, respectively. The fitted parameters are listed in Tables VIII and IX for em and hadronic signal, respectively.

The factor $g_{\mathrm{em},k}$ reflecting the average change of em signal due to the change of $X_{\mathrm{max}}$ scale for a mix of primary species $i$ with relative fractions $f_i$ in zenith-angle bin $k$ is calculated as

$$g_{\mathrm{em},k} = \sum_i f_i \frac{S_{\mathrm{em}}(\langle t \rangle_k - \Delta X_{\mathrm{max}})}{S_{\mathrm{em}}(\langle t \rangle_k)}, \qquad \text{(B2)}$$

where $\langle t \rangle_k$ is the average measured $X_{\mathrm{atm}}(\theta) - X_{\mathrm{max}}$ in a zenith-angle bin $k$: $\sim$241.2 g/cm$^2$, $\sim$335.5 g/cm$^2$, $\sim$429.4 g/cm$^2$, $\sim$556.1 g/cm$^2$, $\sim$777.6 g/cm$^2$, for respective increasing average values of the zenith angle. The factor $g_{\mathrm{had}}$ for the hadronic signal is obtained as

$$g_{\mathrm{had},k} = \sum_i f_i \frac{S_{\mathrm{had}}(\langle t \rangle_k - \Delta X_{\mathrm{max}})}{S_{\mathrm{had}}(\langle t \rangle_k)} \frac{R_{\mathrm{had}}(\langle t \rangle_k - \Delta X_{\mathrm{max}})}{R_{\mathrm{had}}(\langle t \rangle_k)}, \qquad \text{(B3)}$$

where

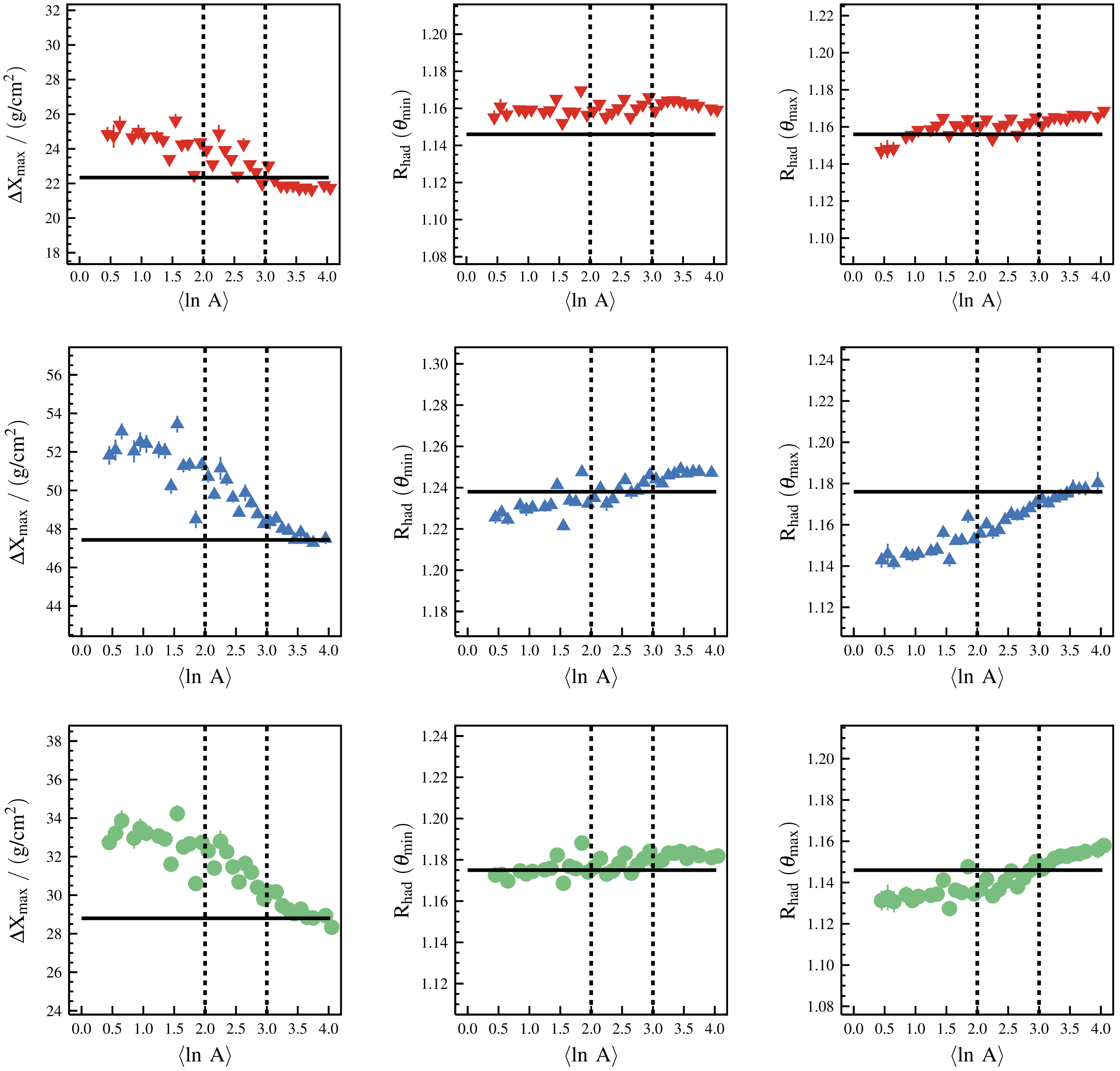

FIG. 12. The values of modification parameters found by the method (points) in the MC samples with artificially modified $\Delta X_{\max}$ and $R_{\rm had}(\theta)$. The values of artificial modifications, shown with solid horizontal lines, are applied to each simulated shower individually. The MC samples contain all possible combinations of primary nuclei with 0.1 steps in relative fractions, iron nuclei are present in all samples (relative fraction ≥0.1), and the mean logarithmic masses of the samples are marked on horizontal axis. Hadronic interaction models used in the tests: Epos-LHC (top row, red), QGSJet-II-04 (middle row, blue), SIBYLL 2.3d (bottom row, green). The systematic uncertainties are taken as the maximum biases within the $\langle \ln A \rangle = 2$ to 3 range indicated with dashed lines.

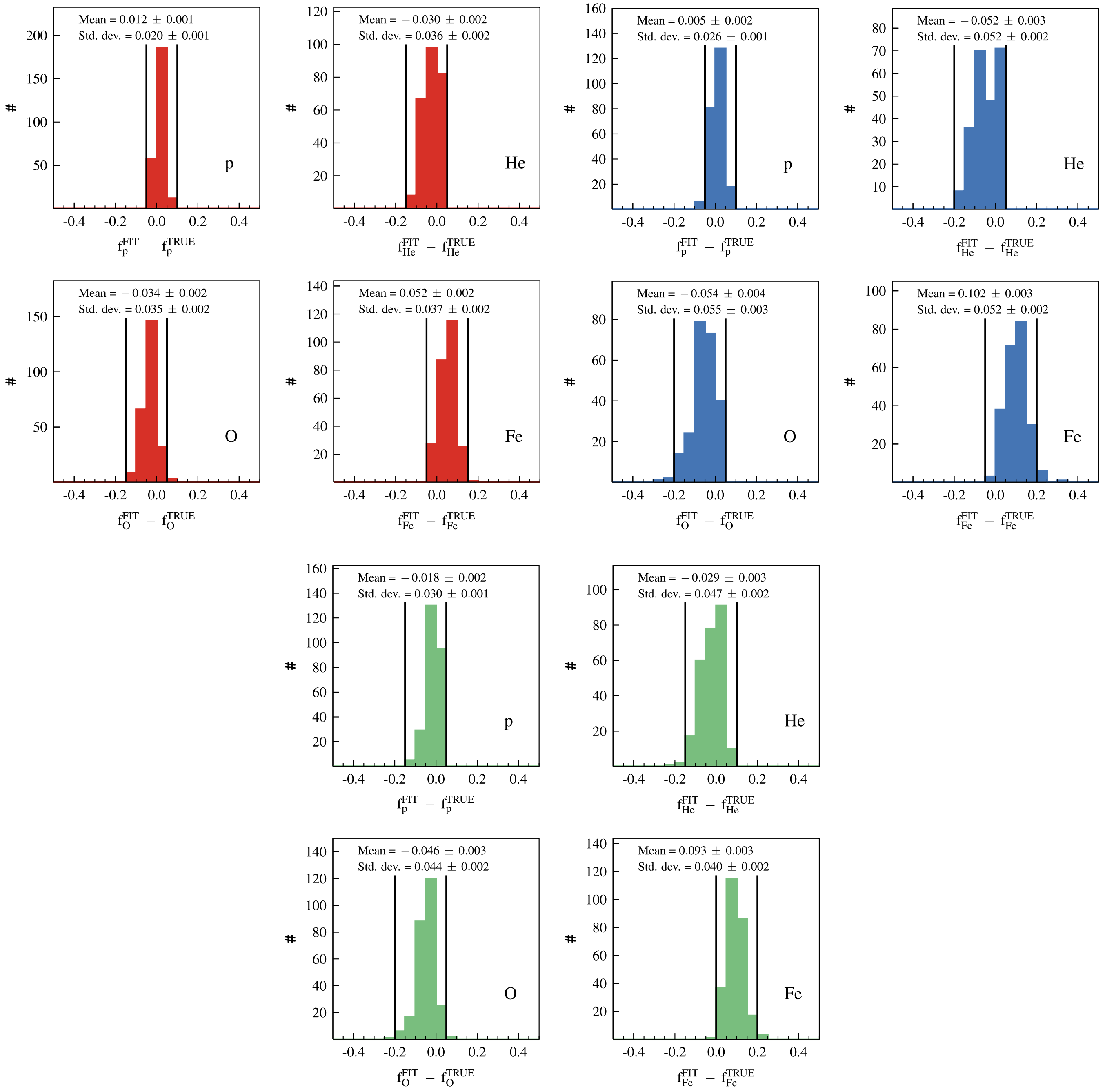

FIG. 13. Biases of the method on the (p, He, O, Fe) fractions for the MC samples with artificially modified $\Delta X_{\mathrm{max}}$ and $R_{\mathrm{had}}(\theta)$ as described in Fig. 12. Additionally, we select the fitted relative fractions with $0.0 \leq f_{\mathrm{p}}^{\mathrm{FIT}} \leq 0.4$, $0.0 \leq f_{\mathrm{He}}^{\mathrm{FIT}} \leq 0.3$, $0.2 \leq f_{\mathrm{O}}^{\mathrm{FIT}} \leq 0.6$, and $0.0 \leq f_{\mathrm{Fe}}^{\mathrm{FIT}} \leq 0.5$ which correspond to the ranges obtained by the data fits. The range considered as a contribution to the systematic uncertainty is indicated by vertical lines. Hadronic interaction models used in the tests; Epos-LHC (top left, red), QGSJet-II-04 (top-right, blue), SIBYLL 2.3d (bottom, green).

$$R_{\text{had}}(\theta) \equiv R_{\text{had}}(\langle t \rangle_k) = R_{\text{had}}(\theta_{\text{min}}) + \tag{B4}$$

$$+ \, (R_{\text{had}}(\theta_{\text{max}}) - R_{\text{had}}(\theta_{\text{min}})) \frac{\langle t \rangle_k - \langle t \rangle_{\text{min}}}{\langle t \rangle_{\text{max}} - \langle t \rangle_{\text{min}}}, \tag{B5}$$

for $\langle t \rangle_{\text{min}}$ and $\langle t \rangle_{\text{max}}$ corresponding to minimum and maximum zenith-angle bins, respectively.

## Appendix C: MC-MC TESTS OF THE METHOD

To evaluate the precision and reliability of the method, we performed tests on the MC simulations with artificially modified predictions on the event-by-event basis using the shift $\Delta X_{\text{max}}$ and the factors $R_{\text{had}}(\theta_{\text{min}})$, $R_{\text{had}}(\theta_{\text{max}})$ obtained from the fits to the measured data, see Table III.

For each model, all possible mixes of (p, He, O, Fe) with relative fractions changing in 0.1 steps are used to estimate the biases on modification parameters as a function of the mean logarithmic mass of the primary beam $\langle \ln A \rangle$. In the tests, we use only the composition mixes containing primary iron nuclei, as there can be additional biases stemming from the assumption on the presence of different species in the primary beam, see discussion in Sec. IV B. For each composition mix, five different samples of simulated showers are randomly selected with the same number of events as in the data ($N = 2239$) and following the shape of the measured energy spectrum. The biases of the method for each composition mix are calculated as an average over the biases for these five random sets. The results of the MC-MC tests are summarized in Fig. 12. We find that for $2 < \langle \ln A \rangle < 3$, the range of the masses inferred from the data analysis (Sec. III B), the maximum overestimation of the fitted $X_{\text{max}}$ scale with the method is 5 g/cm$^2$, and biases on the fitted $\langle S_{\text{had}} \rangle$ scale are within 3%. These resulting values of these MC–MC tests are considered as systematic uncertainties of the method on the modification parameters, see Fig. 16. The systematic uncertainties of the method on the primary fractions are within ±20%, which corresponds to the maximum bias of the method on the fractions inferred from ∼90% of the studied MC samples, see Fig. 13.

TABLE VIII. Fitted parameters of functional dependence of the average em signal part of $S(1000)$ vs distance of $X_{\text{max}}$ to the ground in atmospheric depth units for air showers initiated by a primary particle $i$, see Eq. (B1).

| | $i$ | $S^0_{\text{em}}$/VEM | $t_{\text{em}}/(\text{g/cm}^2)$ | $u_{\text{em}}/(\text{g/cm}^2)$ | $v_{\text{em}}/(\text{g/cm}^2)$ | $w_{\text{em}}$/VEM |
|---|---|---|---|---|---|---|
| Epos-LHC | p | 7.58 | −1,638 | 172 | 42 | $1 \times 10^{-1}$ |
| Epos-LHC | He | 7.76 | −580 | 179 | 81 | $6 \times 10^{-2}$ |
| Epos-LHC | O | 7.83 | −1,412 | 161 | 48 | $9 \times 10^{-2}$ |
| Epos-LHC | Fe | 7.64 | −407 | 204 | 86 | $7 \times 10^{-2}$ |
| QGSJet-II-04 | p | 7.56 | −2,006 | 175 | 35 | $1 \times 10^{-1}$ |
| QGSJet-II-04 | He | 7.71 | −495 | 183 | 87 | $3 \times 10^{-2}$ |
| QGSJet-II-04 | O | 7.77 | −763 | 167 | 72 | $7 \times 10^{-2}$ |
| QGSJet-II-04 | Fe | 7.52 | −278 | 206 | 103 | $2 \times 10^{-3}$ |
| SIBYLL 2.3d | p | 7.57 | −1,152 | 176 | 53 | $1 \times 10^{-1}$ |
| SIBYLL 2.3d | He | 7.62 | −490 | 185 | 88 | $3 \times 10^{-2}$ |
| SIBYLL 2.3d | O | 7.54 | −665 | 187 | 73 | $6 \times 10^{-2}$ |
| SIBYLL 2.3d | Fe | 7.74 | −1,892 | 152 | 39 | $9 \times 10^{-2}$ |

TABLE IX. Same as in Table VIII, but for the average hadronic-signal part of $S(1000)$.

| | $i$ | $S^0_{\text{had}}$/VEM | $t_{\text{had}}/(\text{g/cm}^2)$ | $u_{\text{had}}/(\text{g/cm}^2)$ | $v_{\text{had}}/(\text{g/cm}^2)$ | $w_{\text{had}}$/VEM |
|---|---|---|---|---|---|---|
| Epos-LHC | p | 5.6 | −9,999 | 312 | 9 | 8.3 |
| Epos-LHC | He | 7.2 | −9,419 | 232 | 16 | 8.4 |
| Epos-LHC | O | 9.0 | −1,298 | 228 | 98 | 8.7 |
| Epos-LHC | Fe | 10.7 | −8,304 | 171 | 24 | 9.6 |
| QGSJet-II-04 | p | 5.1 | −1,880 | 314 | 47 | 7.2 |
| QGSJet-II-04 | He | 7.2 | −300 | 255 | 222 | 6.8 |
| QGSJet-II-04 | O | 7.2 | −4,934 | 205 | 34 | 8.7 |
| QGSJet-II-04 | Fe | 9.6 | −1,614 | 166 | 115 | 8.7 |
| SIBYLL 2.3d | p | 4.8 | −8,519 | 283 | 11 | 8.1 |
| SIBYLL 2.3d | He | 6.7 | −3,557 | 225 | 44 | 8.0 |
| SIBYLL 2.3d | O | 8.4 | −1,521 | 207 | 100 | 8.4 |
| SIBYLL 2.3d | Fe | 8.9 | −5,078 | 194 | 35 | 10.1 |

The performance of the method [20] is compatible with these estimations of systematic uncertainties also for a test when one of the models is used for the analysis of the MC samples prepared with other models. This is in agreement with the expectations that the main differences in MC templates between the models are due to the differences in $X_{\max}$ and $S_{\mathrm{had}}$ scales. Finally, we checked that the number of zenith-angle bins does not affect the results unless it is too small (2 bins are not enough to disentangle the hadronic and em parts of the ground signal) or too large (>10) when the event statistics per zenith-angle bin becomes too low.

## Appendix D: DATA DESCRIPTION USING FITS WITH LESS FREEDOM IN MC TEMPLATES

The description of Auger data using fits without any modification to MC templates and with zenith-angle independent $R_{\mathrm{had}}$ are shown in Figs. 14 and 15, respectively.

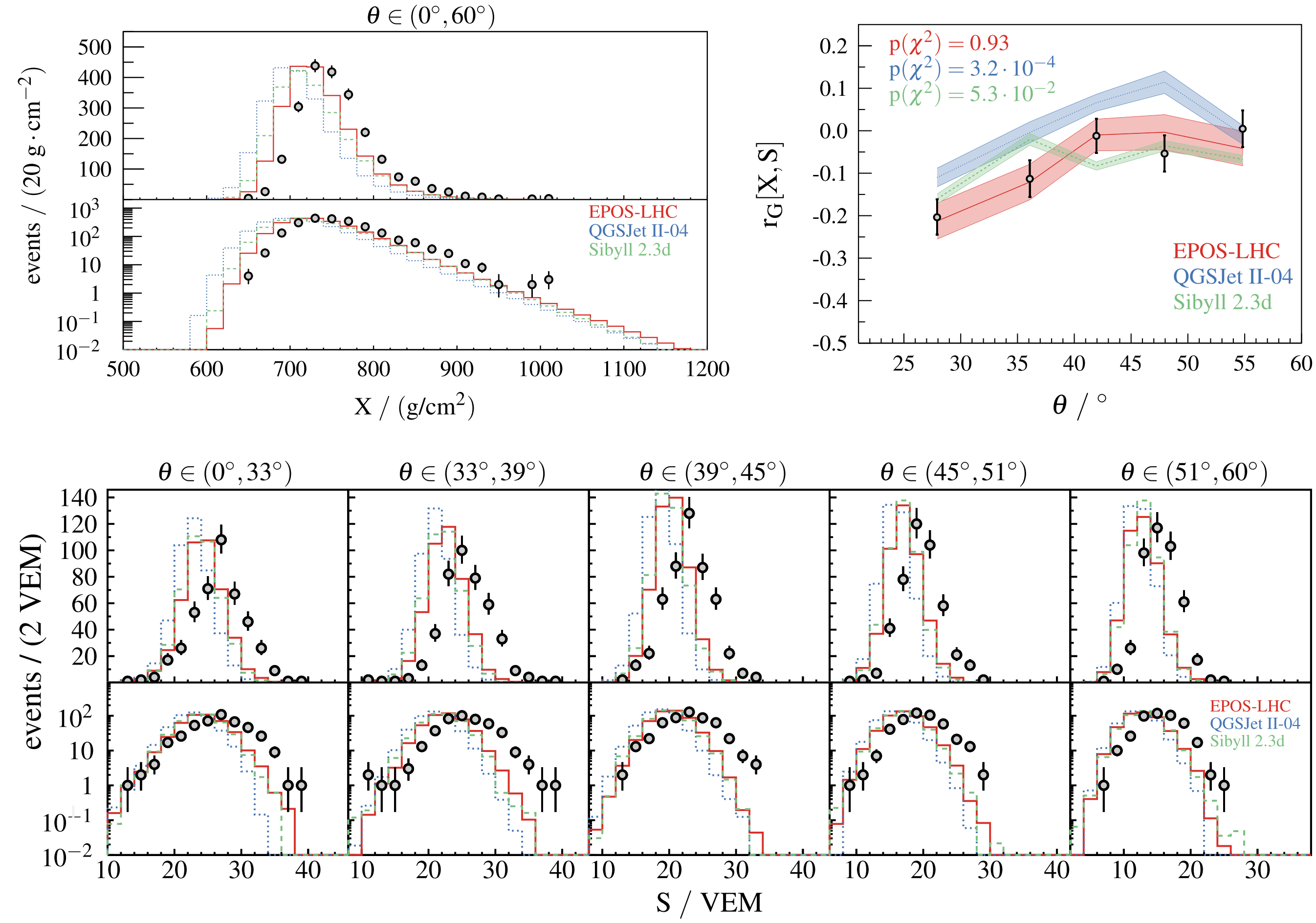

FIG. 14. Same as in Fig. 5, but for the data fits performed without any modification of the predictions of models.

## Appendix E: SYSTEMATIC UNCERTAINTIES

The individual contributions to the total systematic uncertainties are plotted in Fig. 16.

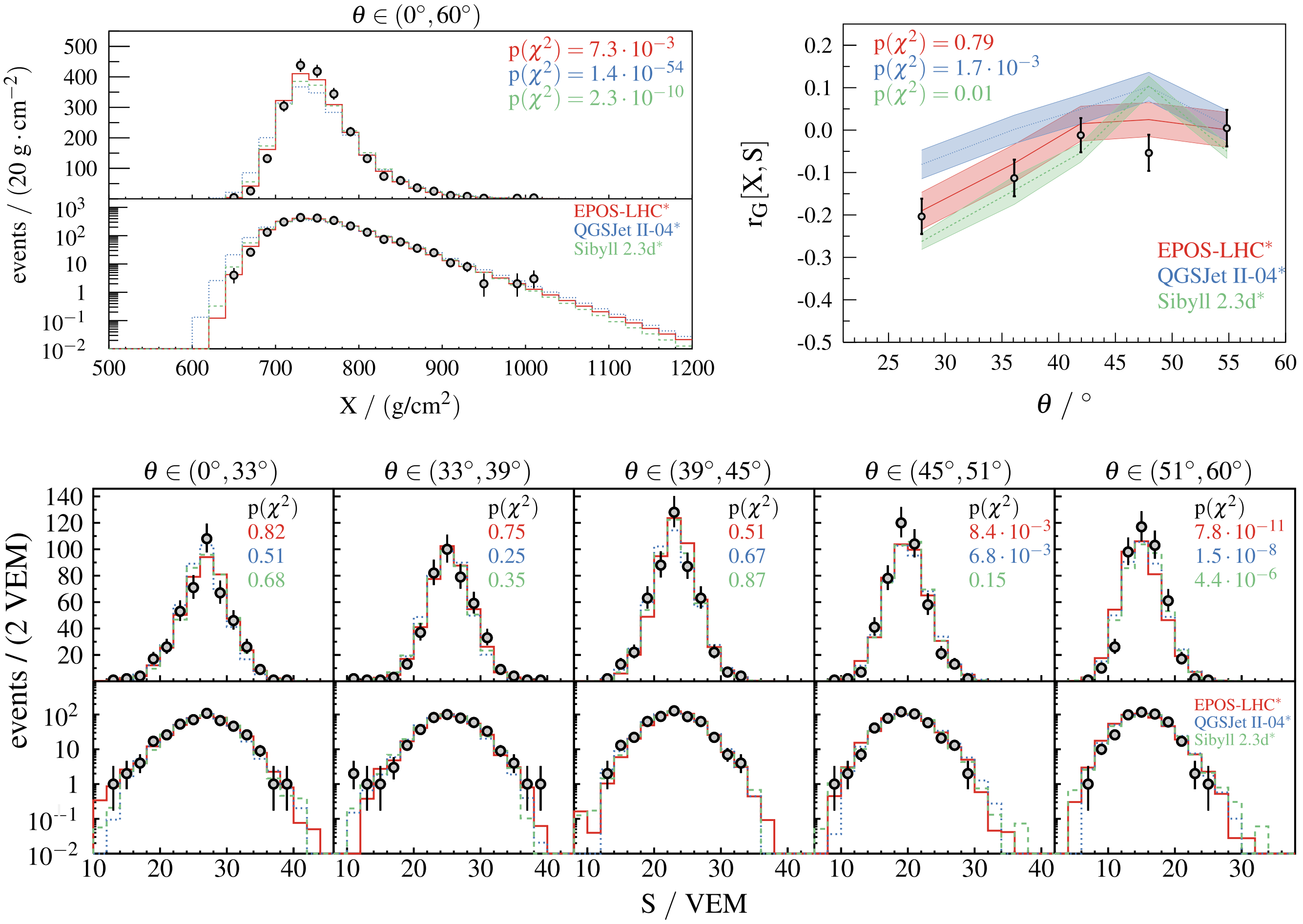

FIG. 15. Same as in Fig. 5, but for the data fits performed with only zenith-angle independent modification $R_{\mathrm{had}}$ of the predictions of models (denoted by $^*$ as templates were modified from the original predictions).

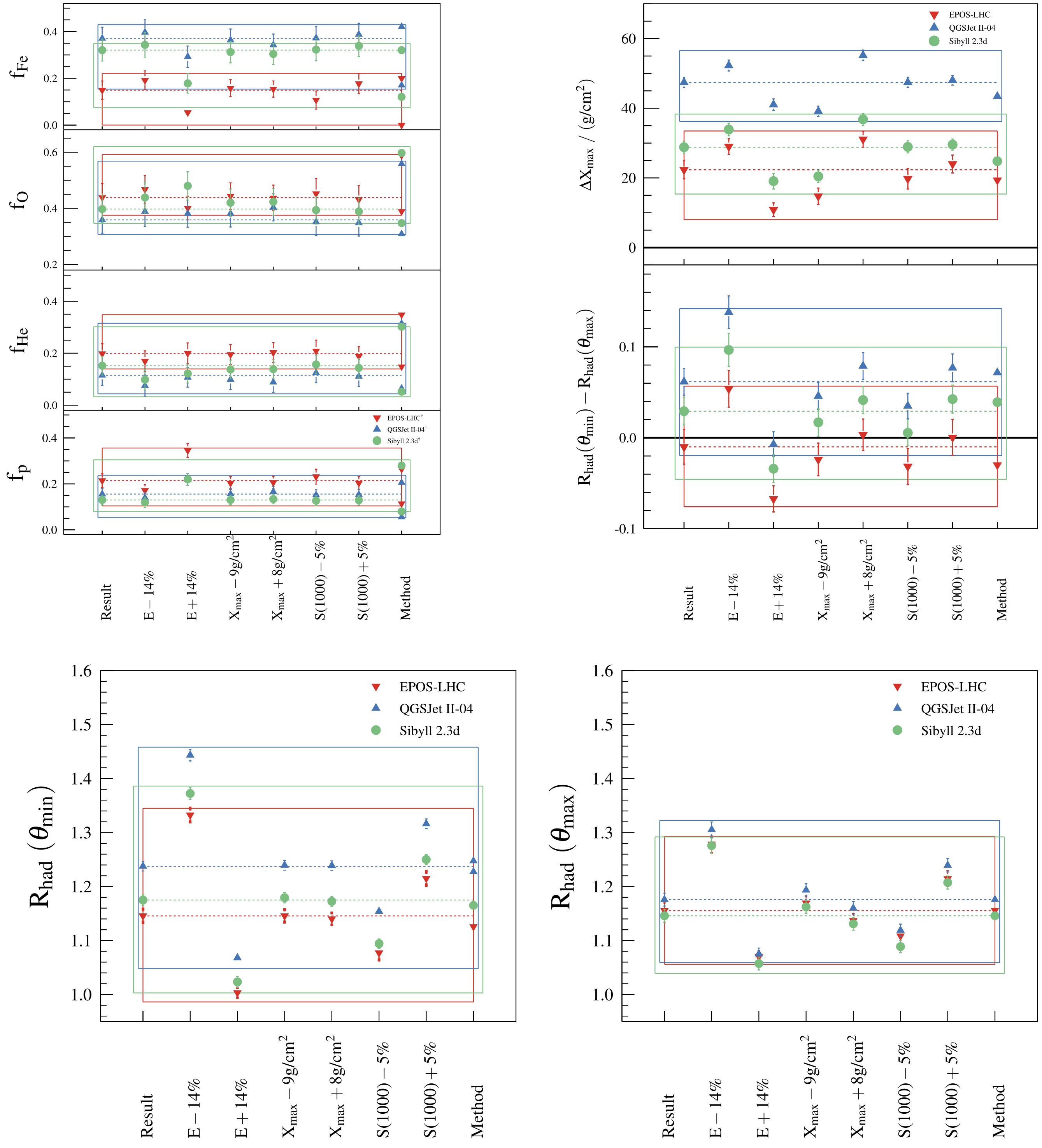

FIG. 16. Contributions of individual experimental systematic uncertainties on energy, $X_{\mathrm{max}}$, $S(1000)$ and of the method biases inferred from the MC–MC tests (see Appendix B) to the total systematic uncertainties on fractions of primary nuclei (top left), $\Delta X_{\mathrm{max}}$ (top right), $R_{\mathrm{had}}(\theta)$ (bottom) and zenith-angle difference of $R_{\mathrm{had}}$ (top right). Colored bands are the total systematic uncertainties obtained by summing individual contributions in quadrature with the best-fit results indicated by dashed horizontal lines.

## Appendix F: SCAN FOR THE LINEAR COMBINATIONS OF EXPERIMENTAL SYSTEMATIC UNCERTAINTIES MOST FAVORABLE FOR THE MODELS

In Fig. 17, the histogram of difference in log-likelihood expressions ($\Delta \ln \mathcal{L}$) for fits using nonmodified and $\Delta X_{\max}$, $R_{\text{had}}(\theta)$ modifications is shown for dense scans in linear combinations of experimental systematic uncertainties. The ranges of these uncertainties were selected in a way to estimate the closest approach to the nonmodified value ($R_{\text{had}}(\theta) = 1,\ \Delta X_{\max} = 0\ \text{g/cm}^2$) even for cases when the uncertainties were out of the range of uncertainties quoted by Auger. It was not possible to find such a linear combination of experimental systematic uncertainties that would decrease the significance of improvement in data description below $5\sigma$.

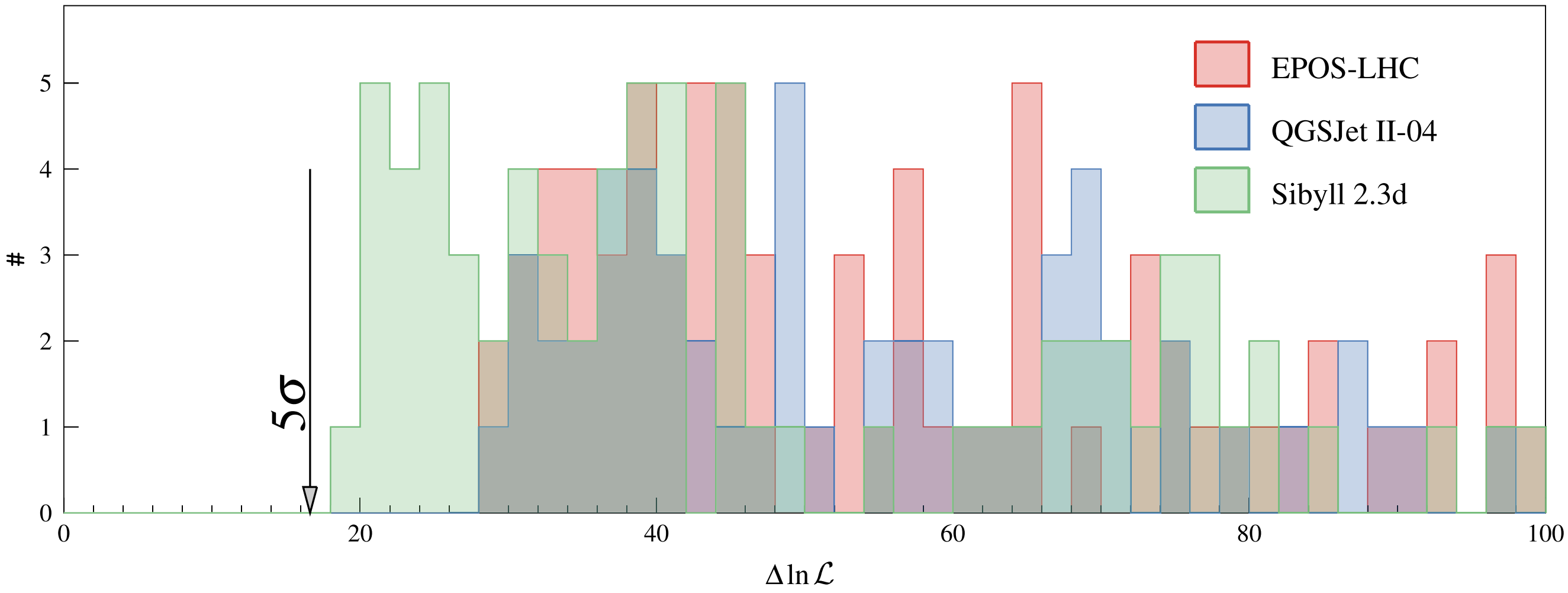

FIG. 17. Number of scans in linear combinations of experimental systematic uncertainties for fits using modifications $\Delta X_{\max}$ and $R_{\text{had}}(\theta)$ to estimate the closest approach to the point ($R_{\text{had}}(\theta) = 1$, $\Delta X_{\max} = 0\ \text{g/cm}^2$) from fits with nonmodified MC templates in the difference of log-likelihood $\Delta \ln \mathcal{L}$. The value estimated using Wilks' theorem in the likelihood-ratio test for the nested model at the level of $5\sigma(\Delta \ln \mathcal{L} \approx 16.62)$ is indicated by the arrow.